\documentclass[
  journal=pasa,
  manuscript=article-type,
  year=2026,
  volume=37,
]{cup-journal}

\usepackage{aas-macros} 
\makeatletter
\renewcommand\@biblabel[1]{[#1]}
\makeatother
\usepackage{amsmath}
\usepackage{amssymb}
\usepackage{xcolor}
\usepackage[nopatch]{microtype}
\usepackage{booktabs}
\usepackage{multirow}
\usepackage{changepage}
\usepackage{ragged2e}
\usepackage{array}
\usepackage{rotating}
\usepackage{lscape}
\usepackage{aas-macros}
\usepackage{siunitx} 

\newcolumntype{L}[1]{>{\RaggedRight\arraybackslash}p{#1}}

\title{From NVSS to RACS: Identifying truly Compact Galactic and Extragalactic sources with Steep Spectra}

\author{Rajat Shinde}
\affiliation{Indian Institute of Science Education and Research -- Pune, 411008, India}

\author{Yogesh Maan}
\affiliation{National Centre for Radio Astrophysics (NCRA -- TIFR), Pune -- 411007, India}
\email[YM]{ymaan@ncra.tifr.res.in}

\author{Apurba Bera}
\affiliation{International Centre for Radio Astronomy Research, Curtin University, Bentley, WA 6102, Australia}
\alsoaffiliation{ASTRON, the Netherlands Institute for Radio Astronomy, Oude Hoogeveensedijk 4,7991 PD Dwingeloo, The Netherlands}


\begin{document}

\begin{abstract}
Compact, steep-spectrum radio sources are key tracers of exotic astrophysical objects such as pulsars and high-redshift radio galaxies. All-sky radio surveys at different frequencies, like the TIFR-GMRT Sky Survey (TGSS) and the NRAO VLA Sky Survey (NVSS), have been usually exploited to identify such tracers. The more recent imaging survey, Rapid ASKAP Continuum Survey (RACS), with higher angular resolution and better sensitivity offers an avenue for a far better identification and characterisation of compact, steep-spectrum sources. In this work, using publicly available RACS images at 887\,MHz and 1.4\,GHz, we present an image-domain characterisation of 171 compact source candidates between declinations $-40^{\circ}$ and $+41^{\circ}$ that were detected and appeared compact at 147\,MHz in TGSS but not detected at 1.4\,GHz in NVSS. Our detailed characterisation resulted in the identification of 66 compact sources, 87 non-compact, diffuse or resolved sources, and 18 sources that are not detected in either of the RACS or NVSS images, implying spectral indices steeper than $-2.0$. Out of the 66 compact sources, 38 have spectral indices steeper than $-1.5$. We demonstrate that a large fraction of the sources in our sample were earlier not detected and resulted in incorrect spectral index limits due to poor imaging quality of NVSS in the Galactic plane. We present the spectral indices and morphological classification of all the sources in our sample and discuss their usefulness in identifying and studying interesting sources such as radio pulsars, high-redshift radio galaxies, and other extragalactic sources.
\end{abstract}

\section{Introduction}

The radio sky is full of sources powered by non-thermal emission. Many of these radio sources exhibit steep radio spectra, i.e., their emission is brighter at lower frequencies. Compact radio sources with steep spectra, although rare, are particularly interesting as they serve as promising candidates for exotic objects like radio pulsars and High-redshift Radio Galaxies (HzRGs).
\par
The frequency dependence of flux density, $S_\nu$, of the steep spectrum radio sources is often characterised as a power-law in the form of $S_\nu \propto \nu^\alpha$, where $\nu$ is the frequency and the power-law spectral index $\alpha$ characterises the steepness of the spectrum.
Pulsars are known for their notable steep spectra with a mean spectral index of $-1.4\pm1.0$ at a population level \citep{BLV13}, but a large fraction of radio pulsars exhibit ultra-steep spectra \citep[with $\alpha<-2.0$;][]{atnf}. HzRGs also exhibit steep spectra, in fact, some of the HzRGs have been identified based on their exceptionally steep spectra \citep[e.g.,][]{DeBreuck00}. In large scale radio surveys with moderate angular resolutions, both pulsars and HzRGs are observed as continuum point sources. While pulsars are truly point sources, HzRGs can often be resolved at arcsec or higher angular resolution.
\par
Searching through the alternative data release of TIFR-GMRT Sky Survey \citep[TGSS;][]{Intema17}, \citet{Frail16} detected continuum radio emission from nearly 300 known pulsars at 147.5 MHz. Similar earlier searches using the NRAO VLA Sky Survey \cite[NVSS;][]{Condon98} had also yielded many known pulsars (\cite{Kaplan1998NVSS}, 79 pulsars, and \cite{Han1999SNR}, 97 pulsars). While majority of the pulsars have been discovered by systematic, untargeted searches for pulsations, image-based targeted search for pulsars in steep spectrum and compact sources can be particularly effective. The first millisecond pulsar (MSP), PSR~B1937+21, was discovered because of the unusually high steepness of its counterpart continuum source 4C~21.53W \citep{Backer82}. Early discovery of several other pulsars was also aided and motivated by their steep spectra in the continuum images \citep[see, e.g.,][]{Lyne87,Hamilton85,Navarro95,Marthi11}. More recently, a search for radio pulsations from 16 promising candidates, selected based on their steep radio spectra, compactness and coincident sky positions with Fermi-LAT unassociated sources, yielded 6 millisecond pulsars and one normal pulsar \citep{Frail2017ImageSearch}. Likewise, in a search for pulsars in candidates identified as compact, steep sources within a small region near the Galactic centre, \citet{Bhakta17} discovered a Galactic disc recycled pulsar with a spectral index of $-2.55$.
\par
 There have also been large scale image-based targeted searches without much success. Searches for pulsations by \citet{Maan18} in 44 steep-spectrum pulsar candidates selected from a spectral index catalogue of 80 \% of the sky \citep{Tiwari19,deGasperin18} did not yield any new pulsars. This search was partly motivated by \cite{deGasperin18} confirmation of an intriguing excess of the compact sources with steep spectra ($\alpha<-1.5$) in the Galactic plane, first indicated by \citet{DeBreuck00}. Several possibilities for the reasons behind no pulsation being detected were discussed, the primary suspect being anomalously high scatter broadening of pulse profiles. \cite{Hyman2021} found two very steep ($\alpha\sim-3$), compact sources near the central bulge of the Galaxy. Higher frequency searches did not yield any pulsations, but these are hypothesised to be most likely scatter-broadened pulsars. Other examples include \citet{Damico85} and \citet{Kaplan00}, with the latter suggesting that most of their compact sources with steep-spectra were HzRGs.
 
 \par
 Despite the inherent difficulties in confirming the pulsars in steep-spectrum sources (scatter-broadening could dominate at lower frequencies and sources would be intrinsically faint at higher frequencies), it is likely that large fractions of sources in the above discussed sample were affected by the limitations of then available surveys. For example, TGSS and NVSS have relatively coarse angular resolutions of 25`` and 45``, respectively. Moreover, poor quality of NVSS images in the Galactic plane might result in incorrect estimates of the spectral indices. Image-based targeted search for pulsations could be made more worthwhile by including the new imaging surveys with enhanced sensitivities and better angular resolutions, as and when those become available, in the initial sample selection itself. With this motivation, in this work, we use the Rapid ASKAP Continuum Survey (RACS) to probe intermediate frequencies \citep[RACS-low and RACS-mid at 887 and 1367\,MHz, respectively;][]{McConnell2020,racs_paper2,McConnell2020,racs4,racs5}, between that of TGSS and NVSS, and characterise and classify 171 steep-spectrum candidates. We determined the flux densities and compactness for these sources using data from the higher-resolution ASKAP images at both frequencies, whenever available, and characterise their spectra using the measurements and upper limits from TGSS and NVSS. Our characterisation and classification of these sources can help in finding new radio pulsars, as well as potentially discovering other exotic Galactic sources that might mimic pulsars in their compactness and steep spectra. Beyond galactic populations, our characterisation and classification methods are also well suited for investigating extragalactic sources like AGNs and HzRGs. A fraction of our candidates might also be compact steep spectrum (CSS) radio galaxies --- powerful young radio galaxies with projected linear sizes $\lesssim 20$ kpc and spectral indices $\alpha \lesssim -0.5$ \citep{ODeaSaikia21}.  While early identified samples were bright (> 10\ Jy at 178 MHz), modern wide-area deep radio surveys have found CSS sources at mJy brightnesses and below. \citet{HerzogCSS} demonstrated that CSS candidates, such as those within the population of infrared-faint radio sources (IFRS), typically exhibit 1.4 GHz flux densities of a few to a few tens of mJy. \citet{GordonCSS} found that CSS galaxies with enhanced star formation are particularly compact, with linear sizes consistently smaller than 10 kpc.

\par
The rest of the paper is structured as follows. The sample selection is described in Section \ref{Sample Selection}, followed by the methodology in Section \ref{Methodology}. Results are presented in Section \ref{Results}, followed by discussion and concluding remarks in Sections \ref{Discussion} and \ref{Conclusion}.

\section{Sample selection}\label{Sample Selection}
The spectral index catalogue provided by \citet{deGasperin18} includes all the sources that are visible in the TGSS and/or NVSS at 147\,MHz and 1.4\,GHz, respectively. It extends to -40 degrees declination, accounting for 80 percent of the sky. We use this catalogue to select two samples of compact sources with steep-spectra, one in the Galactic plane and another covering the rest of the sky covered by TGSS and NVSS. For our sample selection, we use all the sources which are detected only in TGSS and not in NVSS, and hence, there is only an upper limit on the spectral indices of these sources in the catalogue. Out of the total 1396515 entries in the catalogue, the above sources constitute less than 1 per cent (6386 sources).
\par
Not all the sources with upper limits on their spectral indices are compact. To evaluate their compactness, we define the compactness parameter, $r$, as the ratio of the peak flux density to the total flux density. We use the compactness parameter and the spectral index limits to make our sample selections. Our Galactic plane (GP) sample consists of all the point-like sources (i.e., the sources are modelled well with a single Gaussian component) within the Galactic latitude range of $-4^\circ$ to $+4^\circ$, and which have spectral indices steeper than $-1.5$ and compactness parameter more than 0.9. For a few sources, the compactness parameter was below the above threshold of 0.9. However, the large uncertainty in their compactness parameter implied that the threshold 0.9 was well within 1$\sigma$ of the actual measurement. We also included these sources in the selected sample. This led to the selection of a total of 119 sources. A cross-match with the ATNF pulsar catalogue using \texttt{psrcat}\footnote{\url{http://www.atnf.csiro.au/research/pulsar/psrcat}} \citep{atnf} suggested 12 of these sources to be known pulsars and these were discarded. The GP sample consists of the remaining 107 sources.
\par
The selection of the sources for the off-Galactic plane (oGP) sample followed similar criteria as the GP sample, with two important differences: the sources were outside the Galactic latitude range of $-4^\circ$ to $+4^\circ$, and a spectral index threshold of $-1.8$ was used instead. The primary selection resulted in 126 sources, out of which 11 were found to be known pulsars and discarded. The oGP sample consists of the remaining 115 sources.
\par
Majority of the radio pulsars, one of our primary targets, lie in the Galactic plane. The excess of compact sources with steep-spectra in the Galactic plane, confirmed by \citet{deGasperin18}, further suggest those sources to have the the highest potential to reveal unknown pulsar populations or other exotic Galactic objects. Moreover, given the relatively poor quality of NVSS images in the Galactic plane, the actual uncertainties on the spectral indices in the Galactic plane are expected to be larger. To account for this, and to probe a larger population of Galactic sources, we have used a slightly flatter spectral index threshold of $-1.5$ in our GP sample (than the one used in our oGP sample, $-1.8$). We also note that for a fixed spectral index threshold, the fraction of extra-galactic sources contributing to the sample is expected to be much larger in the off-Galactic plane. Our slightly steeper spectral index threshold of $-1.8$ for the oGP sample helps in minimizing this fraction. Overall, we hope that the above division of the sample in the GP and oGP aids in focusing more on the Galactic sources that are compact and have steep spectra.
\par
Our GP sample includes 29 sources with spectral indices steeper than $-1.8$. These 29 sources, together with the 115 sources in the oGP sample, form an all-sky sample of 144 compact sources with spectral indices steeper than $-1.8$. However, as the ASKAP survey covers the sky till $+41^\circ$ declination, our study effectively focuses on those sources lying in the overlap of the two regions, i.e., from $-40^\circ$ till $+41^\circ$ declination. These are a total of 171 in number, of which 78 belong to GP, including the ones not steeper than $-1.8$ and the remaining 93 are from the oGP sample. 

\begin{figure*}
    \centering
\vspace{1cm}

    \includegraphics[width=0.331\textwidth]{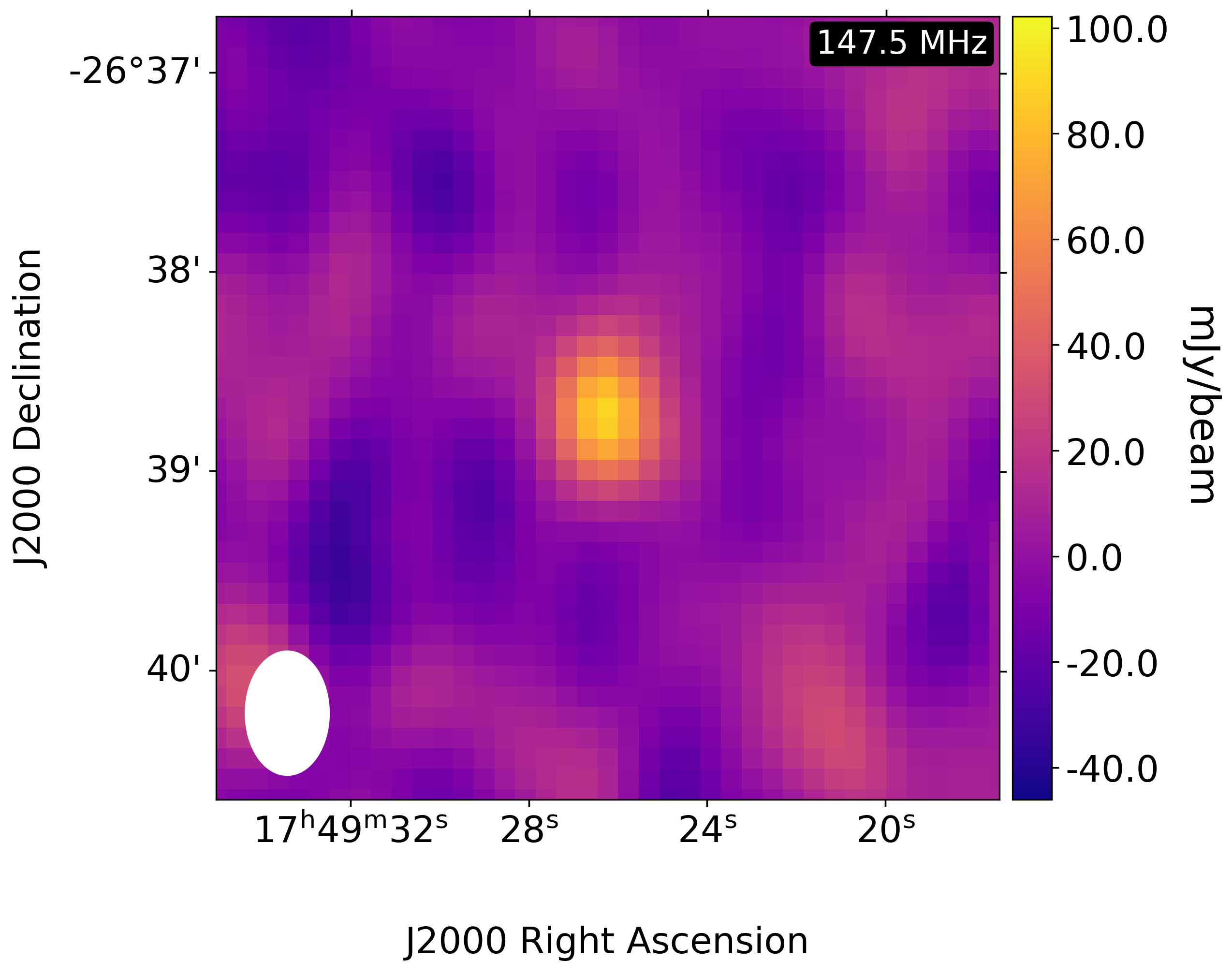} \hfill
    \includegraphics[width=0.330\textwidth]{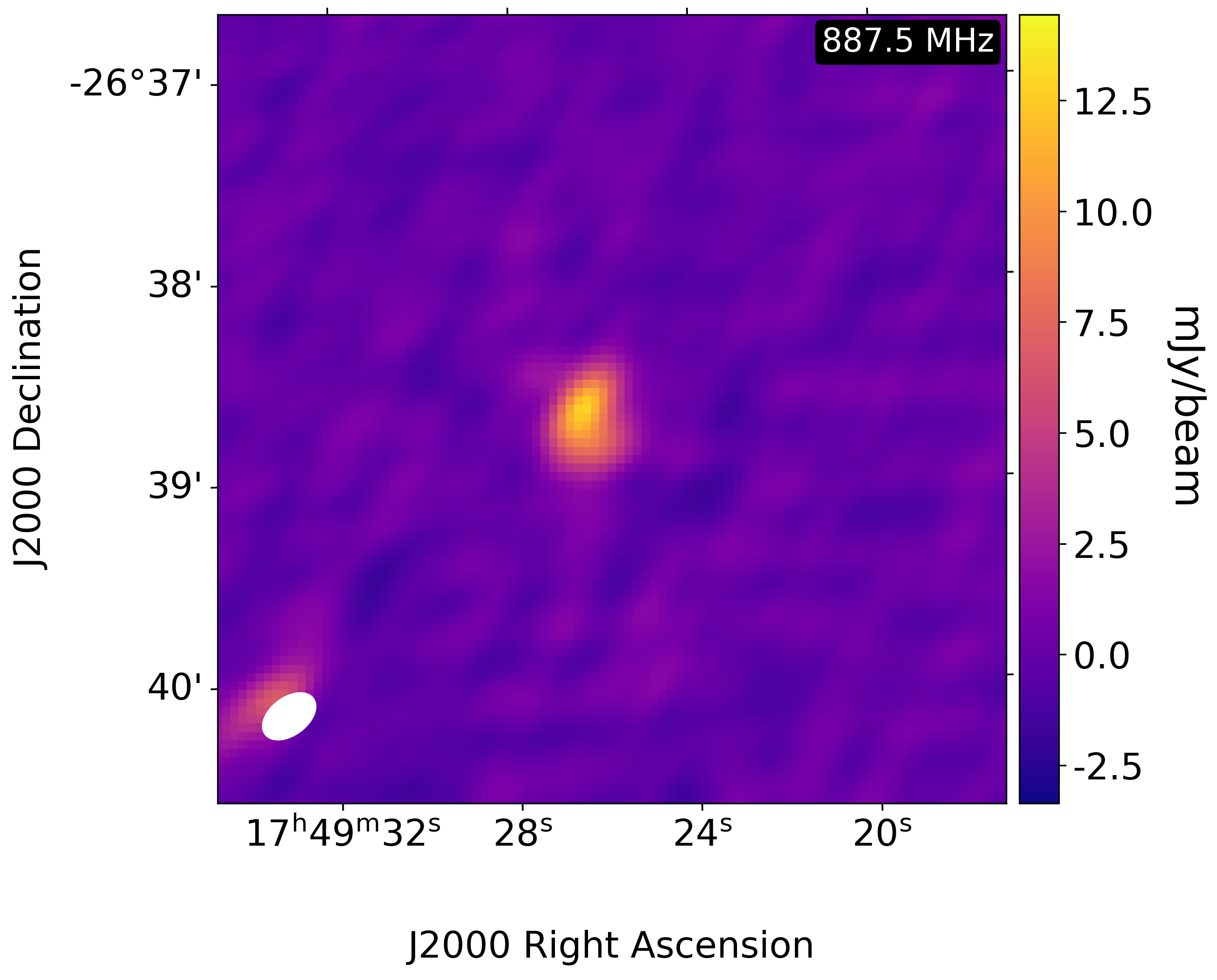} \hfill
    \includegraphics[width=0.331\textwidth]{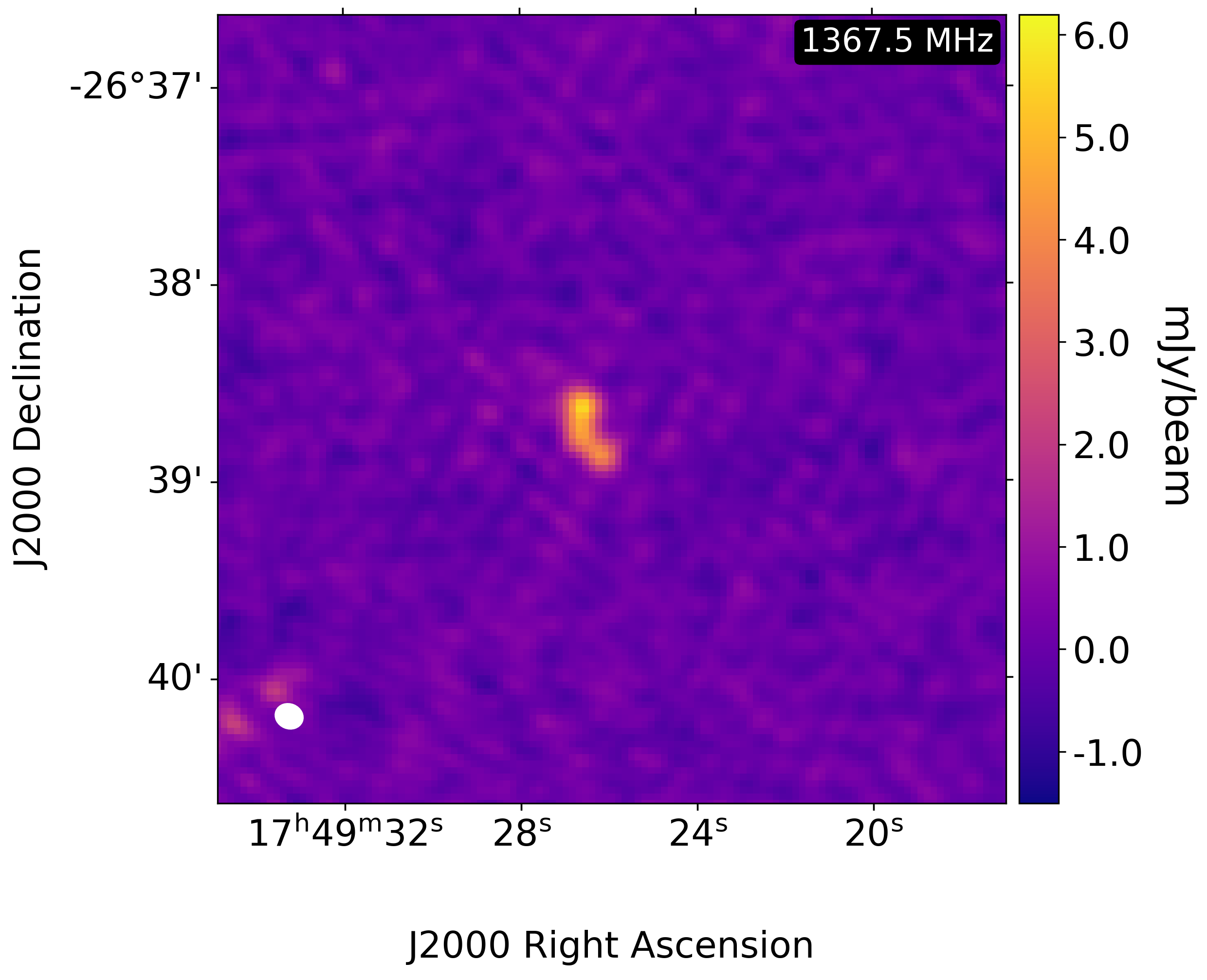}

    \par
    \vspace{1cm}

    \includegraphics[width=0.331\textwidth]{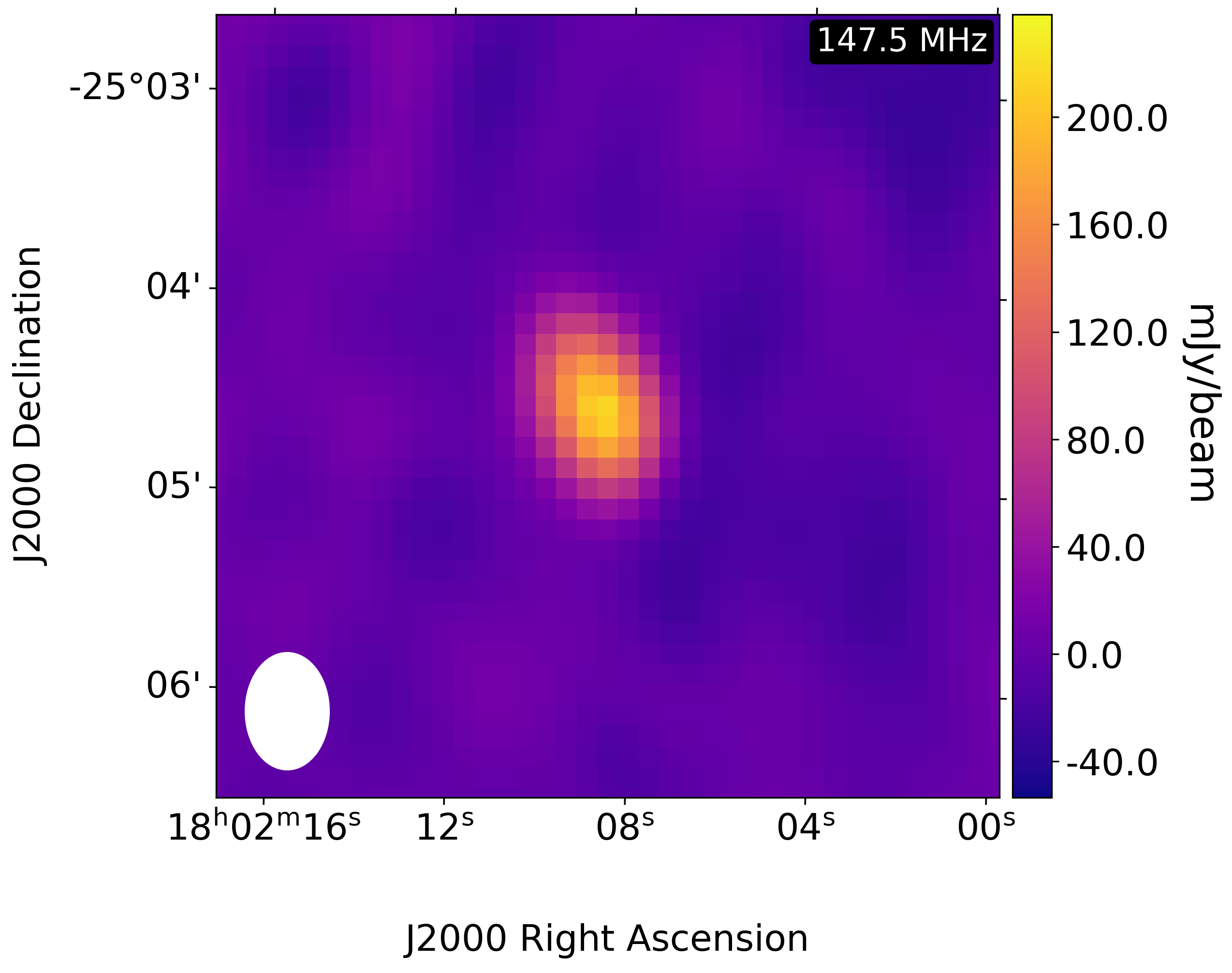} \hfill
    \includegraphics[width=0.330\textwidth]{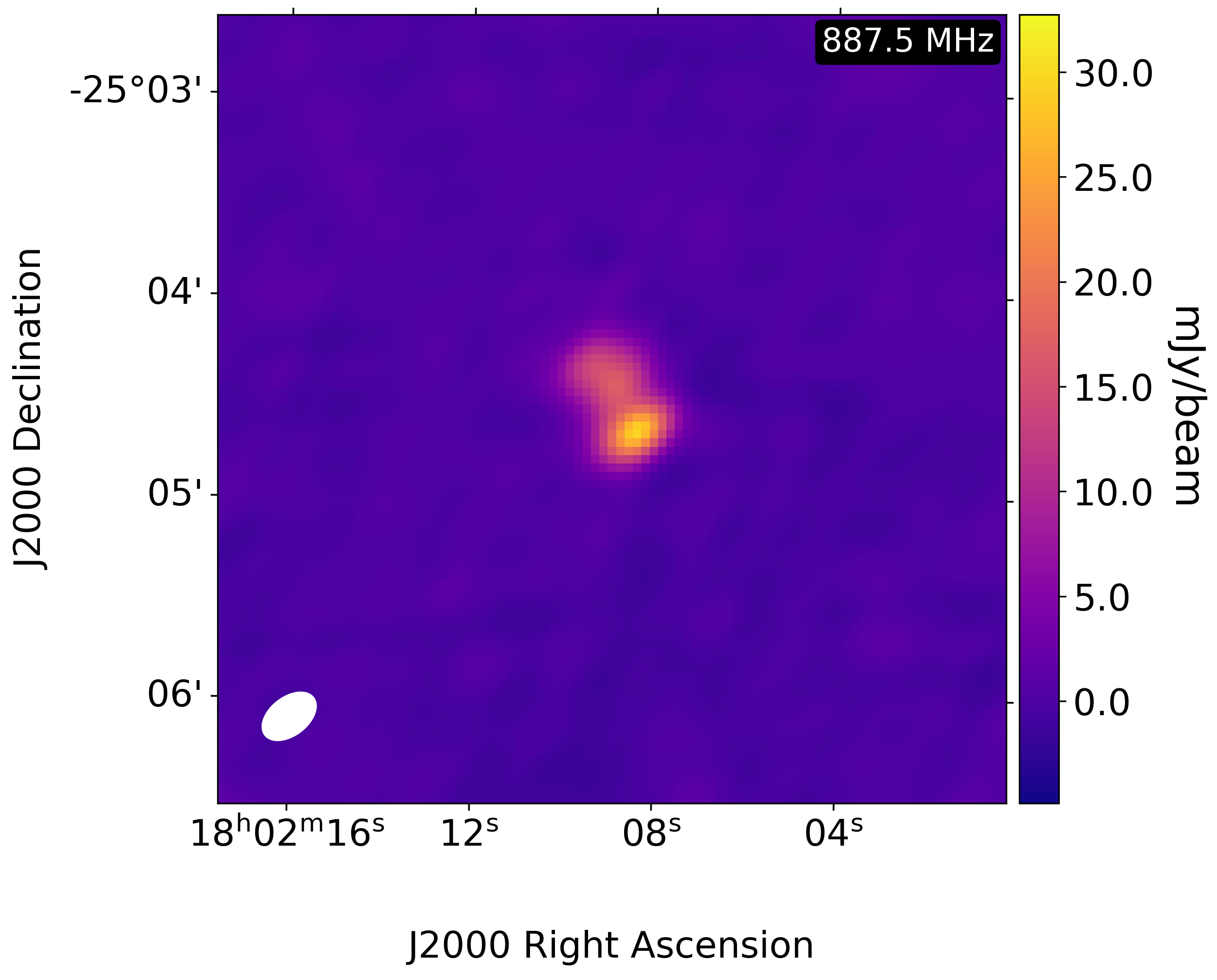} \hfill
    \includegraphics[width=0.331\textwidth]{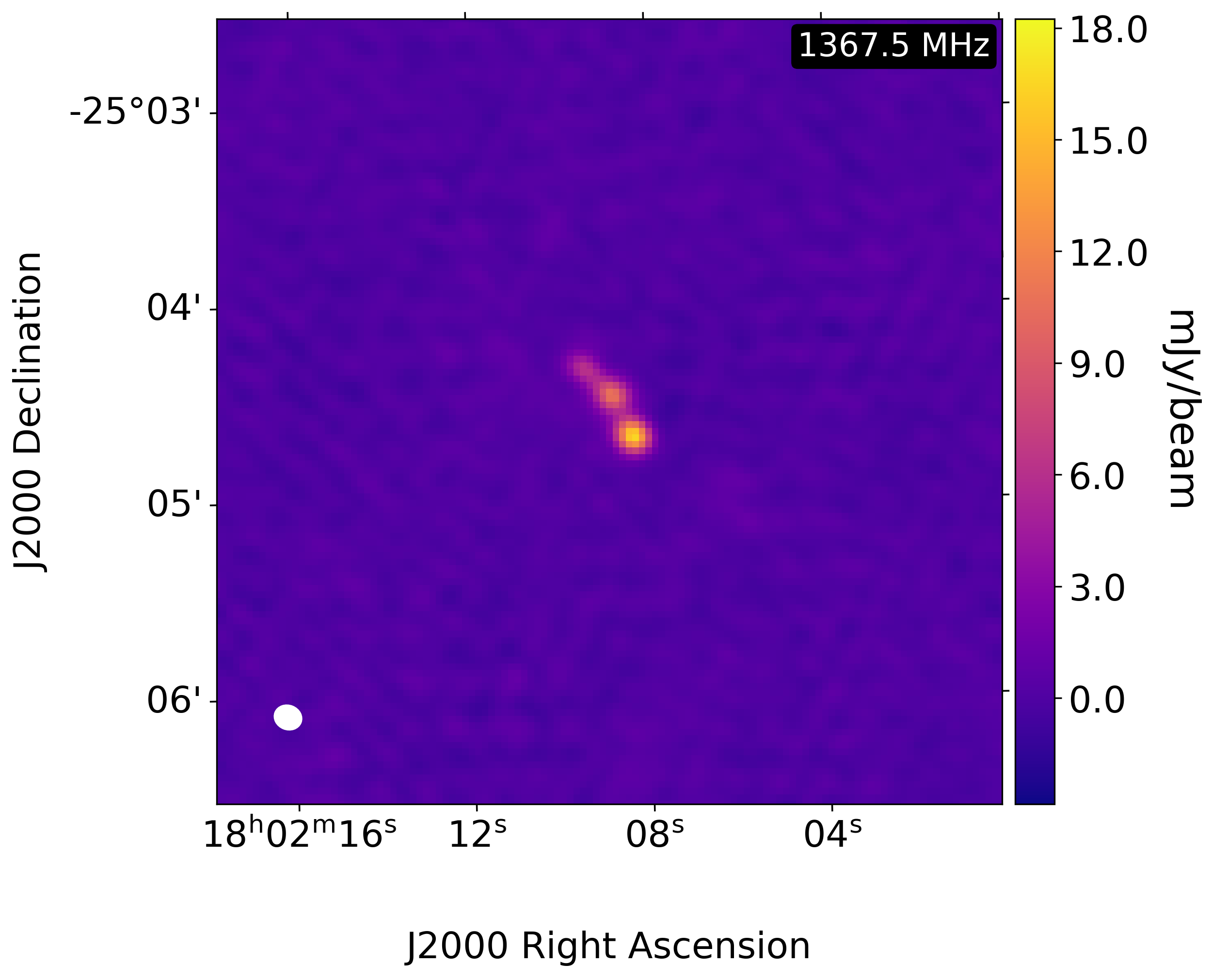}

    \par
    \vspace{1cm}

    \includegraphics[width=0.331\textwidth]{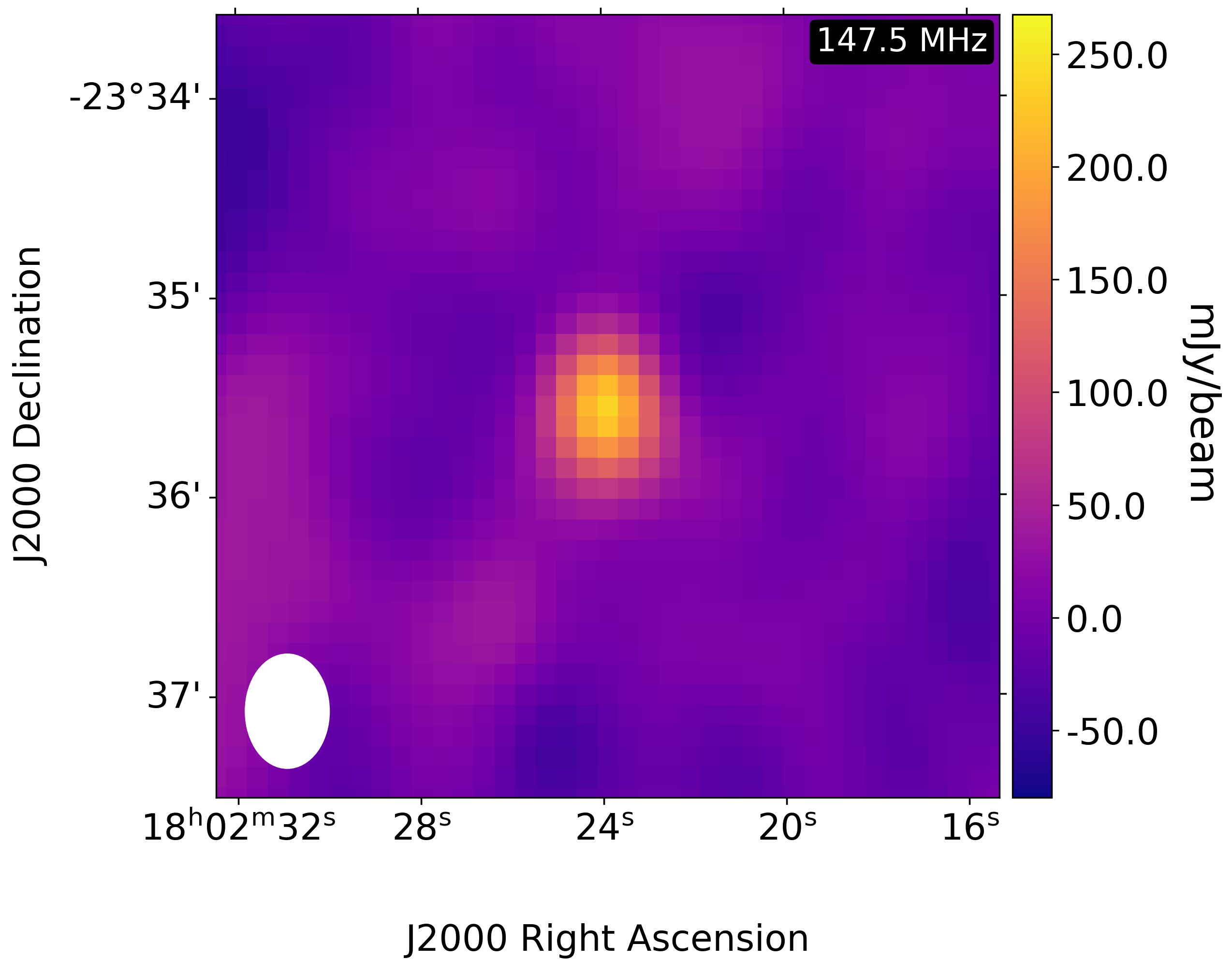} \hfill
    \includegraphics[width=0.330\textwidth]{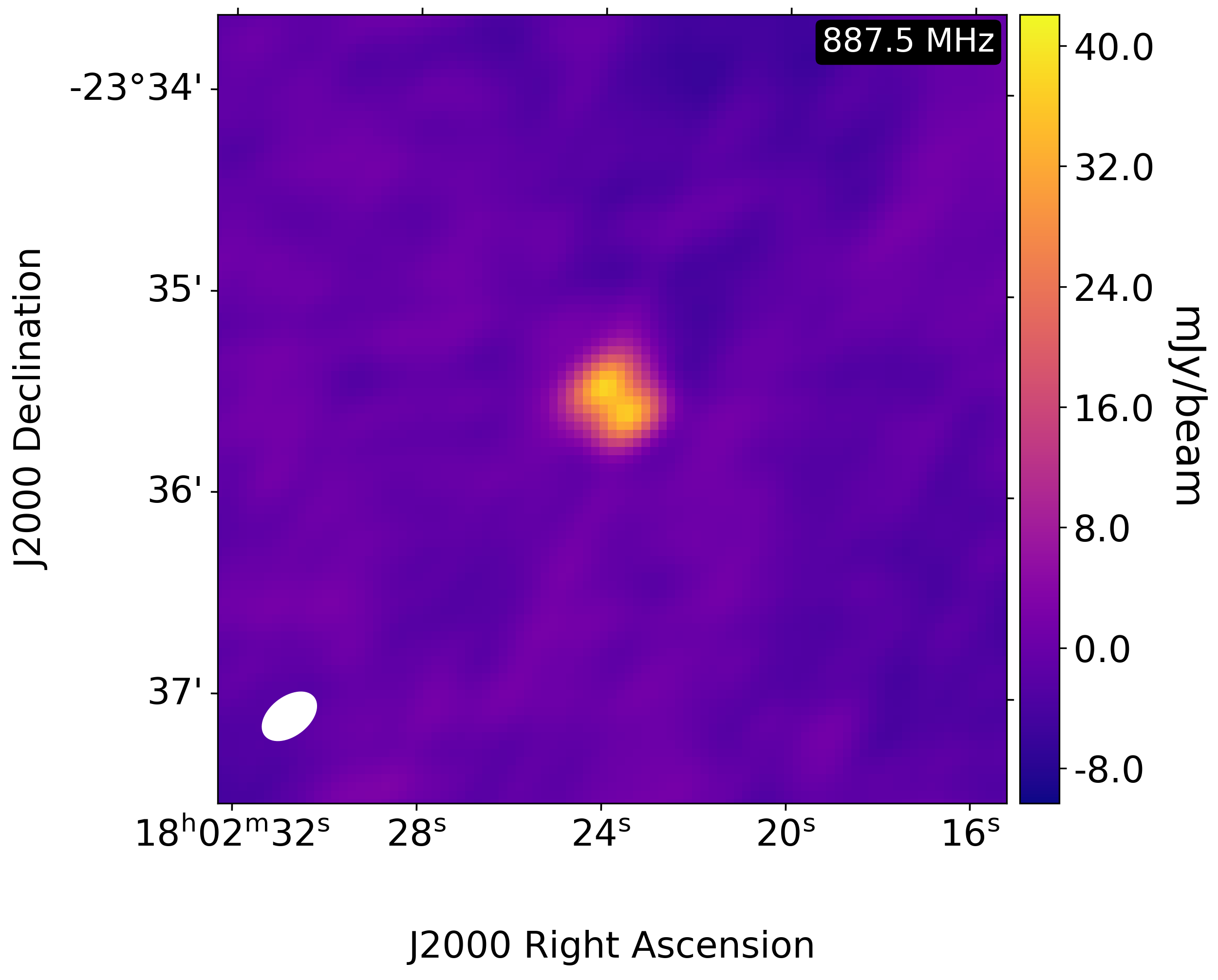} \hfill
    \includegraphics[width=0.331\textwidth]{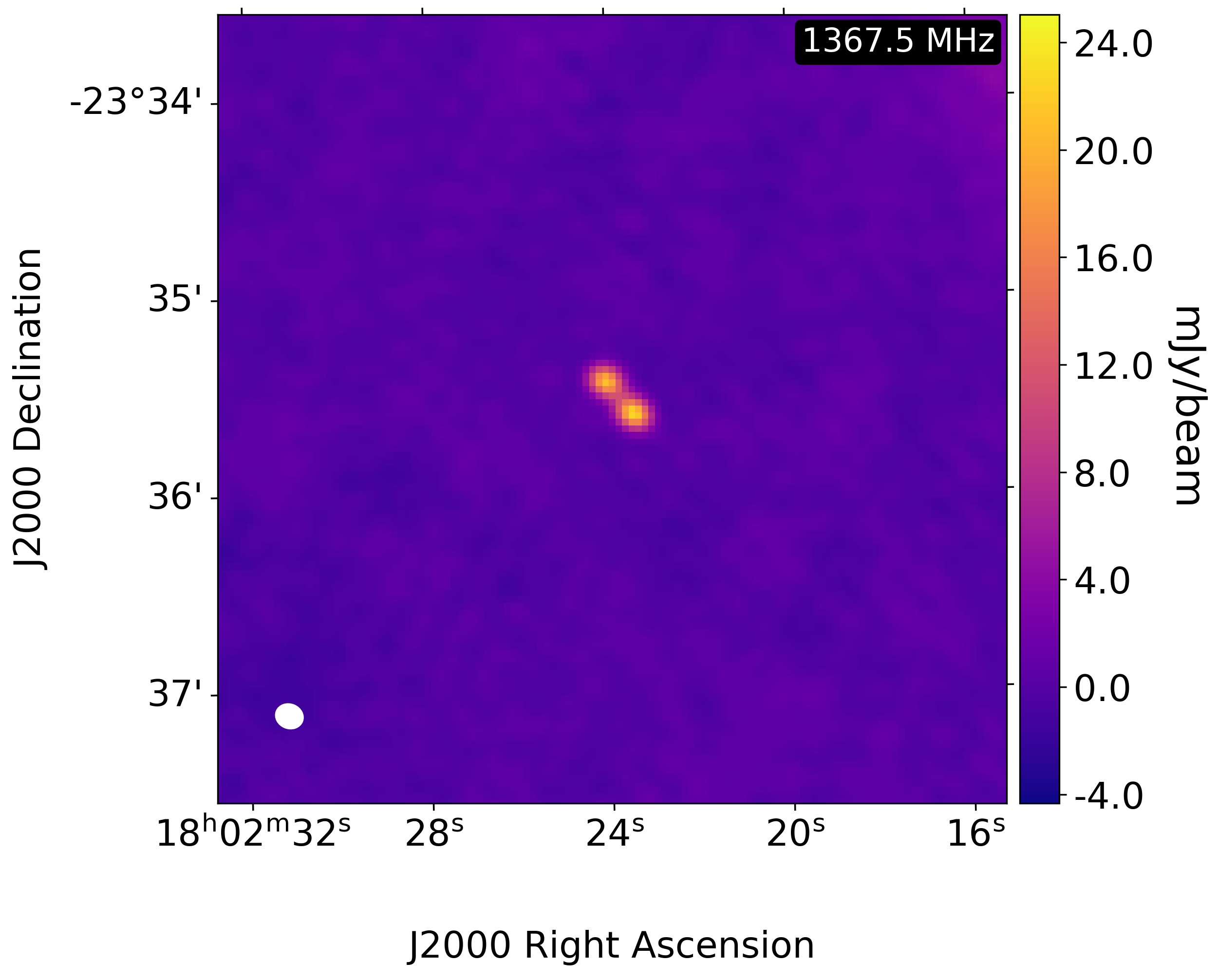}

    \caption{Example of a few sources which appear compact in TGSS (left), but are clearly resolved in RACS-low (centre) and RACS-mid(right). The synthesised beam associated with each image is shown as a white ellipse at the lower left corner.}
    \label{fig: TGSS vs RACS}
\end{figure*}

\section{Methodology}\label{Methodology}
RACS complements and fills a critical niche between the meter-wavelength TGSS and the gigahertz NVSS surveys. Apart from having a substantial sky coverage overlap with the two surveys, RACS observations also offer much better angular resolution and sensitivity \citep{McConnell2020}. The higher angular resolution greatly helps in distinguishing truly compact sources from the ones which appear compact at a lower resolution but are resolved at higher-resolution observations, as illustrated in Figure \ref{fig: TGSS vs RACS}. The data availability at 1367\,MHz further provides an opportunity to study source morphology at a frequency very close to the NVSS frequency, where all the sources in our sample are supposed to be very faint. 
 \par
First, RACS-low and RACS-mid images containing the sky positions of the individual sources from the above described two samples were identified and downloaded from the CSIRO ASKAP Science Data Archive \citep[CASDA\footnote{\url{https://research.csiro.au/casda/}},][]{casda2017}. Individual images were then searched for any continuum emission using the Common Astronomy Software Applications (CASA) package \citep{casa}\footnote{\url{https://casa.nrao.edu/}}. For a few of the sources, we also used images from the ASKAP Variables and Slow Transients \citep[VAST,][]{Murphy2013VAST} and Survey and Monitoring of ASKAP's RFI environment and Trends \citep[SMART,][]{Lourenco2024SMART}). We also note that for some of the sources, images from multiple pointings were available. In such cases, the images where the sources were present closer to the pointing centres and visibly less contaminated from artefacts (e.g., the deconvolution artefacts) were selected for any further analysis. After visual confirmation of the presence of any emission, whenever the emission appeared to be point-like, a circular region was selected around the source and Gaussian fitting was performed with CASA. A source is classified as `detected' if its peak flux density (hereafter, \textit{pfd}) obtained from fitting is more than 4 times the uncertainty on \textit{pfd}, as the \textit{pfd} uncertainty was found to be a good representative of the local rms background ($\sigma_l$). The Gaussian fit also provides parameters such as the integrated flux density (\textit{ifd}), associated uncertainties, more accurate sky position of the source, and source size in terms of the semi-major/minor axes of the fit. We found the compactness, $r$, and the associated uncertainty, $\sigma_r$, for all the point-like sources using the flux density values obtained from RACS-mid images at 1367 MHz: $ r= ({pfd}_{1367\, \text{MHz}})/({ifd}_{1367\, \text{MHz}})$. We classified sources with $r + \sigma_r > 0.9$  as \textit{compact}, and \textit{non-compact} otherwise. Further details on the distribution of compactness values and its role in source classification are presented in the following sections.

\begin{figure*}[t]
\centering

\begin{minipage}{0.32\textwidth}
    \centering
    \includegraphics[width=\linewidth]{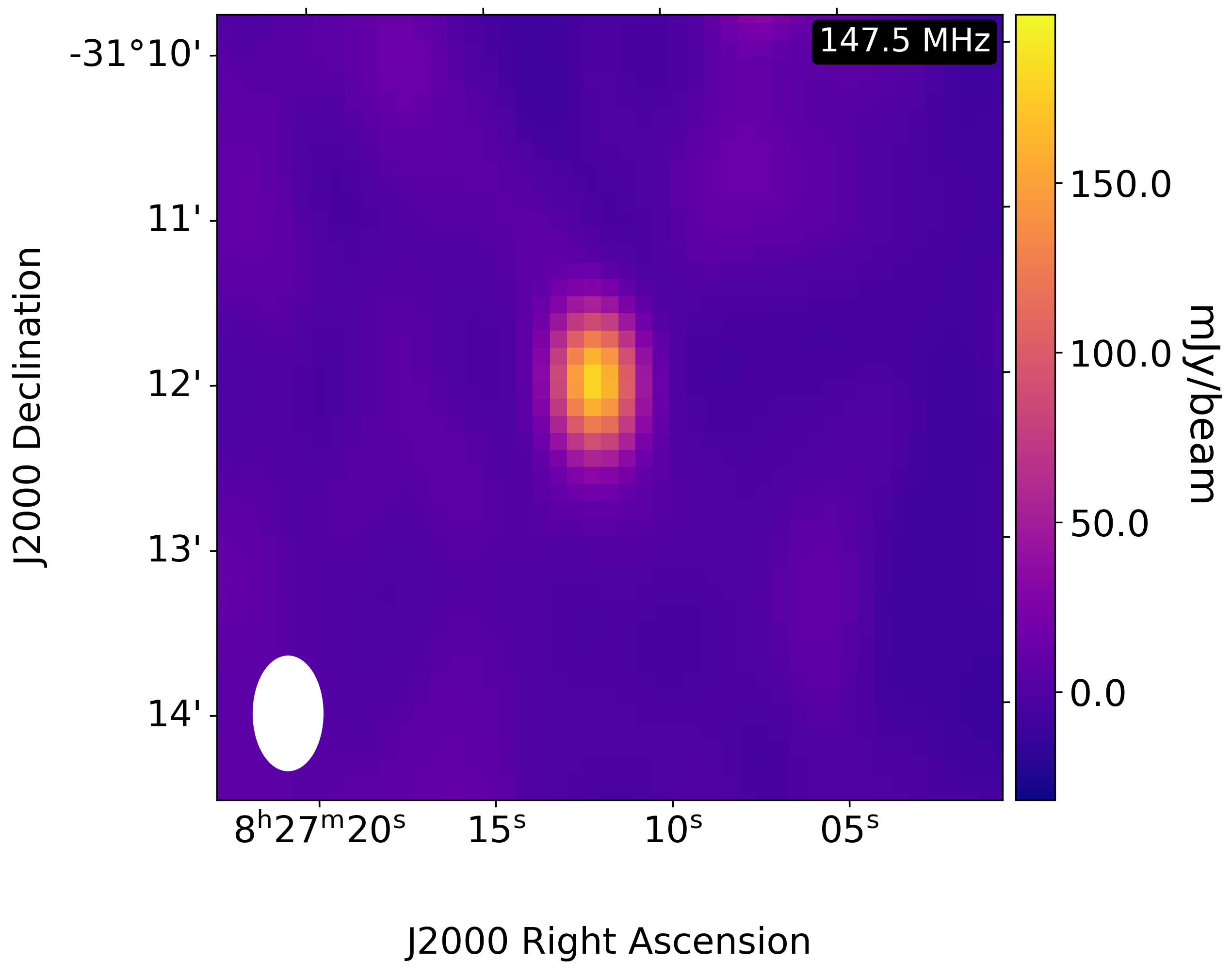}
    \small (a)
\end{minipage}
\hfill
\begin{minipage}{0.32\textwidth}
    \centering
    \includegraphics[width=\linewidth]{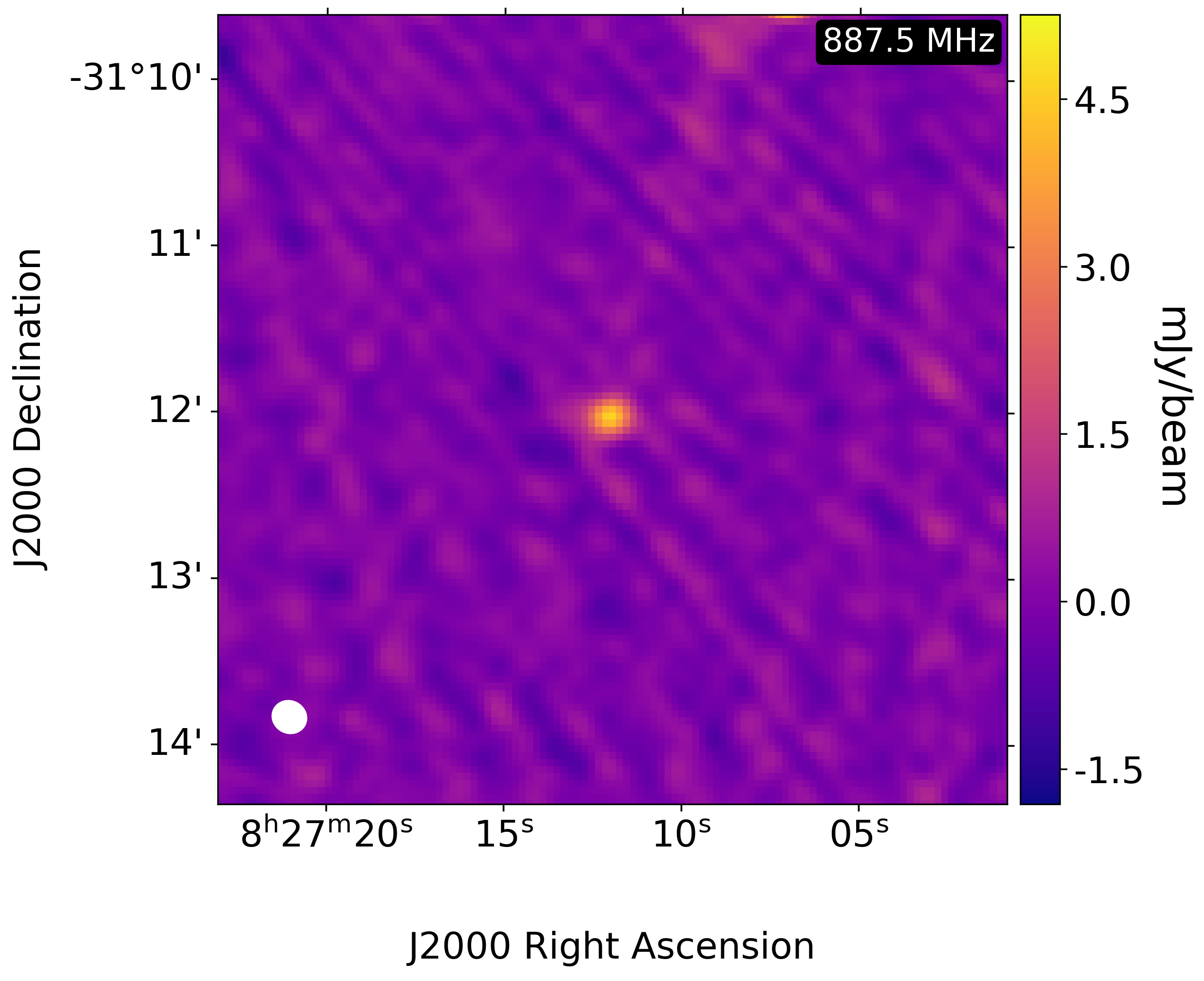}
    \small (b)
\end{minipage}
\hfill
\begin{minipage}{0.32\textwidth}
    \centering
    \includegraphics[width=\linewidth]{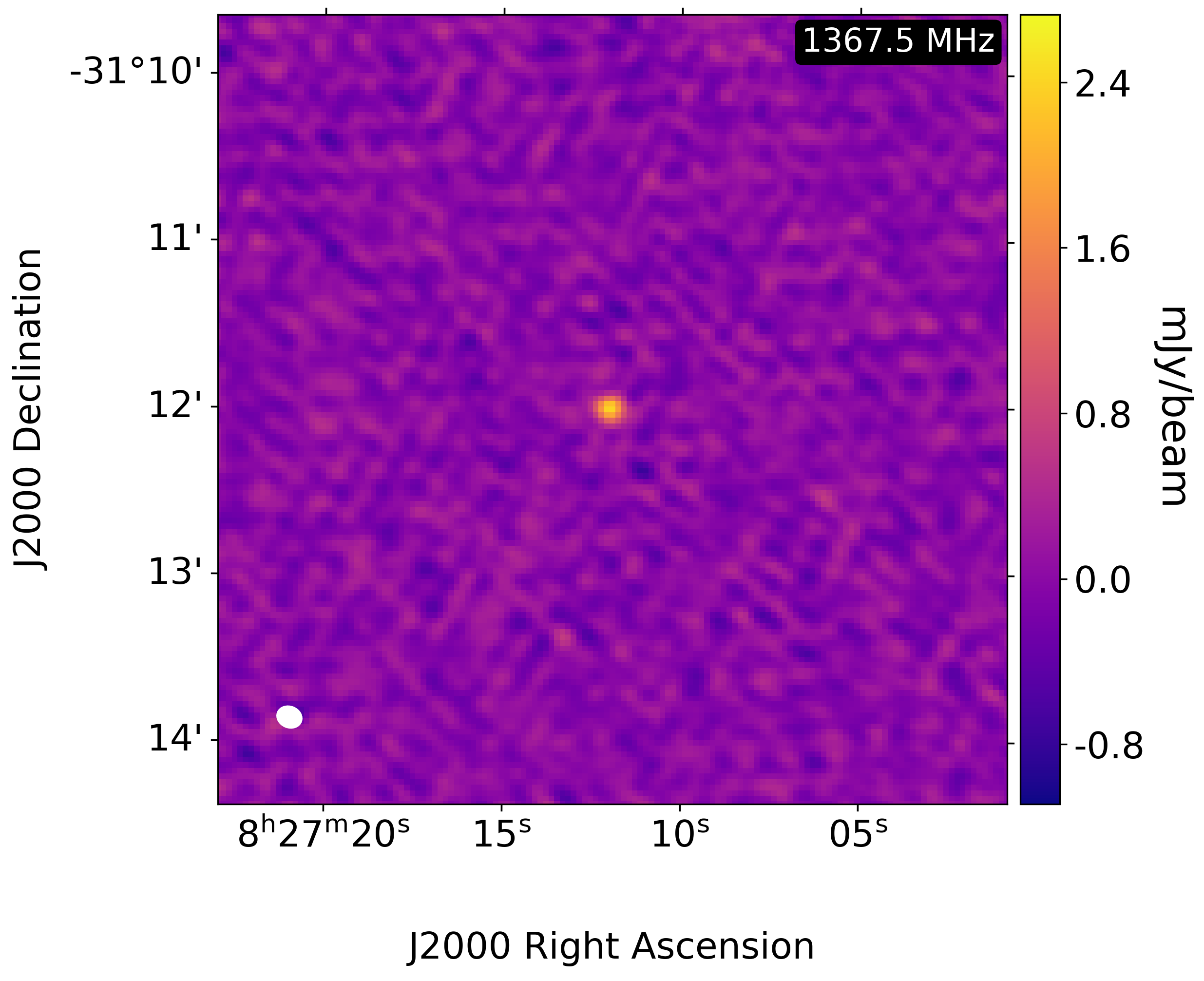}
    \small (c)
\end{minipage}

\vspace{1ex}

\begin{minipage}{0.45\textwidth}
    \centering
    \includegraphics[width=\linewidth]{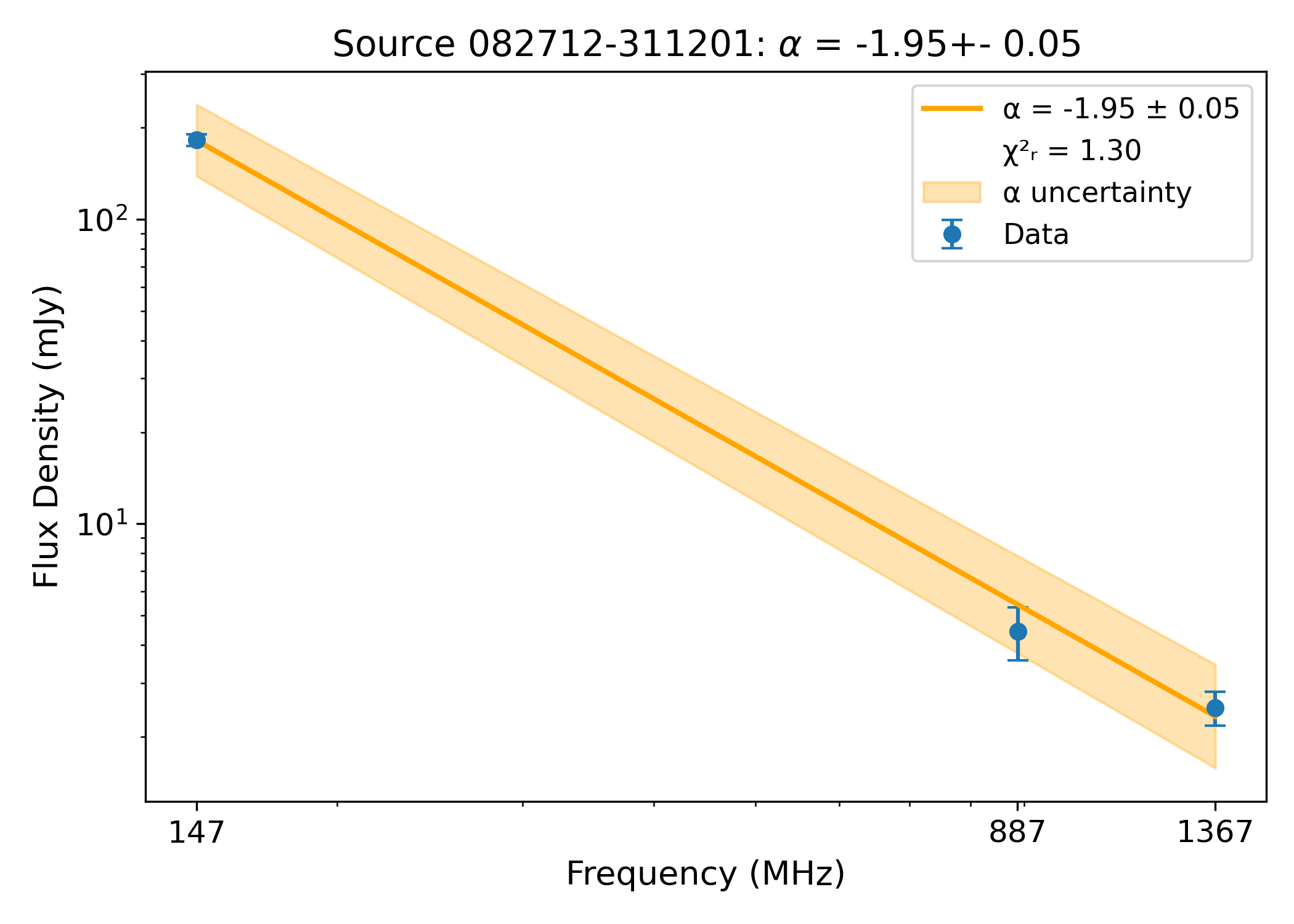}
    \small (d)
\end{minipage}

\caption{Radio images of the source 082712-311201 from the oGP sample, as seen in the TGSS (a), RACS-low (b) and RACS-mid (c) surveys. A power-law fit to the corresponding flux densities is shown in panel (d).}
\label{fig:multi_freq_and_curve_fitting}

\end{figure*}

For the sources that were not detected at 1367\,MHz, their 887\,MHz flux densities were used instead to estimate their compactness. Sources that appeared resolved or extended (i.e., compared to the beam-size) in either of the RACS-low or RACS-mid surveys were classified as \textit{resolved}. For these sources, the FITS images at both frequencies were first convolved to the approximate TGSS resolution of 25"$\times$25", and then the Gaussian fitting procedure was carried out. Similarly, for the sources that were not detected at either of the frequencies, the images were convolved to the approximate TGSS resolution. Whenever this resulted in the detection of a source, it was classified as \textit{diffuse} and the fitting procedure was carried out on the smoothed image. When a source was detected at 887\,MHz but a detection with (\textit{pfd}$>4 *\textit{pfd}_{error}$) at the 1367 MHz became possible only after smoothening, it was also classified as \textit{diffuse}. In such cases, we also convolved and smoothened its 887\,MHz counterpart and noted the new flux densities instead. No compactness measurement was performed for \textit{resolved} and \textit{diffuse} classified sources. Sources without detection at either of the frequencies are referred to as \textit{`non-detection'} or \textit{ND}.

\par
We estimated the new spectral index, $\alpha$, for each of the sources by a linear fit to the three flux density measurements (from TGSS, RACS-low and RACS-mid) in log-log scale. We have used the \textit{pfd} values for the \textit{compact} sources, while \textit{ifd} values were for the \textit{non-compact}, \textit{resolved} or \textit{diffuse} sources. Apart from the measurement uncertainties in the flux densities (\textit{pfd} or \textit{ifd}) obtained from the Gaussian fitting, we also accounted for the systematic uncertainties (by adding them in quadrature) while estimating the spectral indices. The systematic error in RACS-low for a flux density measurement \textit{S} is given by (\cite{McConnell2020}) $ \Delta S= 0.5\,\text{mJy} + 0.07S$. The systematic errors in RACS-mid flux density measurements depend on the relative position of the source within the phased-array feed (PAF) footprint. While for most of the sources the systematic errors are $\approx 6\%$, sources close to the edges of the PAF footprint have up to $14\%$ systematic flux density errors \citep{racs4}. We thus assume a $\approx 10\%$ error on the flux density of all sources as systematic uncertainties, which is a conservative assumption in most cases, for the purpose of estimating $\alpha$. We note that assuming a 14\% systematic uncertainty does not significantly alter our results.
The goodness of fit is estimated in terms of the reduced chi-square, $\chi_r^2$. Figure \ref{fig:multi_freq_and_curve_fitting} shows an example of a compact source (082712-311201) detected in all the three surveys and the corresponding fit for estimating $\alpha$. For the sources that were not detected at 1367\,MHz (RACS-mid), the spectral index is computed simply using the two flux density measurements from TGSS and RACS-low, at 147\,MHz and 887\,MHz, respectively. For the sources classified as ND, we used the upper limits on the flux densities as three times the rms (i.e., $3\sigma_l$) value estimated around the source positions at the respective frequencies. Then, we computed the two-point spectral index limits (147\,MHz- 887\,MHz and 147\,MHz- 1367\,MHz) using these upper limits, and chose the steeper of the two estimates. The spectral index limit thus obtained is an upper limit on $\alpha$, i.e., $\alpha$ could be lower than this limit, and hence, the intrinsic spectrum could be much steeper.

\par
Tables~\ref{tab:table1} and~\ref{tab:table2} summarise the estimated spectral indices, compactnesses ($r$), their associated uncertainties, reduced chi-square ($\chi_{r}^2$) of the fits, and the final classifications of the sources, for the GP and oGP samples. Also shown are the original upper limits ($\alpha^{\mathrm{NVSS}}_{\mathrm{TGSS}}$) on the spectral indices \citep[from][]{deGasperin18} obtained using TGSS flux densities and NVSS upper limits. Finally we classify sources as \textit{ultra-steep} when they have $\alpha\leq-1.8$. Candidates for radio pulsars and HzRGs, for example, can be identified as compact sources with steep ($-1.8 < \alpha < -1.5$) or \textit{ultra-steep} spectra. 

\section{Results}\label{Results}
In total, 143 sources have detections at all the three frequencies, and their $\alpha$ values have been estimated using power-law fits. For 118 of these, the power-law fits are reasonably good, with $\chi_r^2$ value below 2.5. A total of 18 sources remain undetected at either of the RACS frequencies, and thus classified as \textit{ND}. This generally imposes an ultra-steep spectral index limit for these sources, using $3\sigma_l$ limits on their flux densities, as described in Section \ref{Methodology}. The remaining 10 sources are detected at 887\,MHz but could not be detected at 1367\,MHz even after smoothing the images to match with the TGSS angular resolution. We directly computed the 147--887\,MHz two-point $\alpha$ for these 10 sources.
\par
Of the 78 sources in the GP sample, we find 31 to be compact. Likewise, of the 93 sources in the oGP sample, 35 were classified as compact. Across the full sample of 171 sources, 46 have their $\alpha \leq -1.8$. A further 6 are within a sigma error of -1.8. These 52 sources, irrespective of their morphological classification, are classified as \textit{ultra-steep}. All of the 18 \textit{ND} sources, unsurprisingly, are classified as \textit{ultra-steep}, with half of them having spectral indices steeper than -2.5.
The remaining 34 ultra-steep sources comprise 19 compact sources (with 4 in the Galactic plane), 11 non-compact and 4 diffuse sources. Similarly, we note that 19 of 48 sources with steep spectra ($<-1.8 < \alpha < -1.5$) are found to be compact, including 4 sources having $\alpha$ consistent with $-1.5$ within one sigma error.

\par
For completeness, we mention that our total sample of 171 sources is classified into 29 \textit{resolved}, 6 \textit{diffuse}, 52 \textit{non-compact} sources, along with the 18 \textit{ND} and 66 \textit{compact}. The \textit{ND} and \textit{diffuse} classifications exhibit the steepest spectral indices, with averages of -2.47 and -1.82, respectively. The 10 sources detected at 887 MHz, but not-detected at 1367 MHz, include 5 classified as \textit{compact}, 3 as \textit{non-compact}, 1 \textit{resolved} and 1 as \textit{diffuse}. Their average spectral index is -2.1. We recall that the sources in our sample are steep-spectrum sources by design. The average $\alpha$ for the whole sample (excluding \textit{ND}s) is -1.42, with the source with the flattest spectrum having an index of -0.57. The average $\alpha$ in the GP sample is -1.16, flatter than that of -1.66 in the oGP sample. 

\section{Discussion}\label{Discussion}

\begin{figure}
    \centering
    
        \centering
        \includegraphics[scale=0.35]{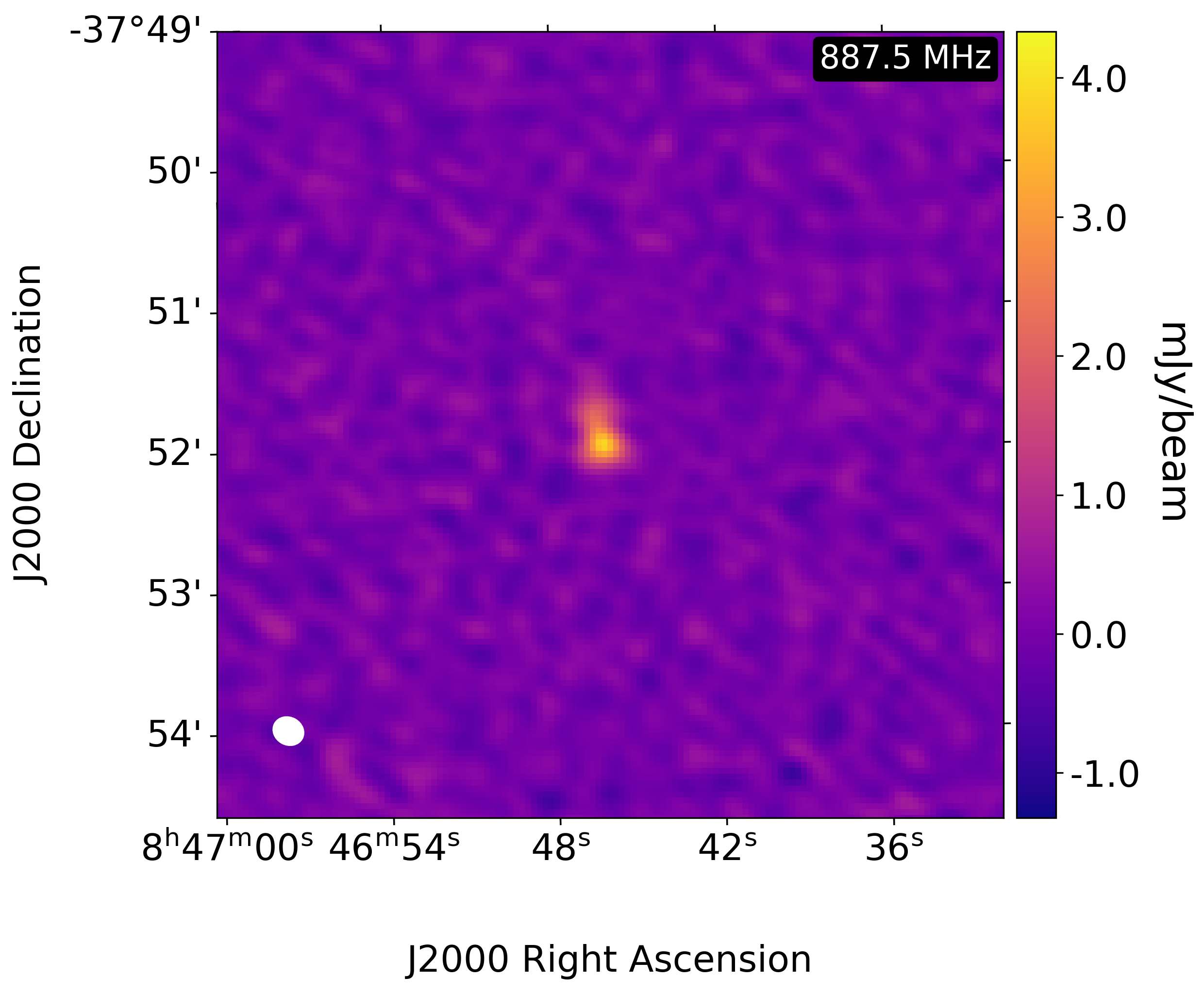}
        \caption{An example of a source classified as \textit{resolved} owing to a tiny extension towards higher declinations.}
        \label{fig:3}
\end{figure}

\begin{figure}
    \centering
    
        \centering
        \includegraphics[scale=0.64]{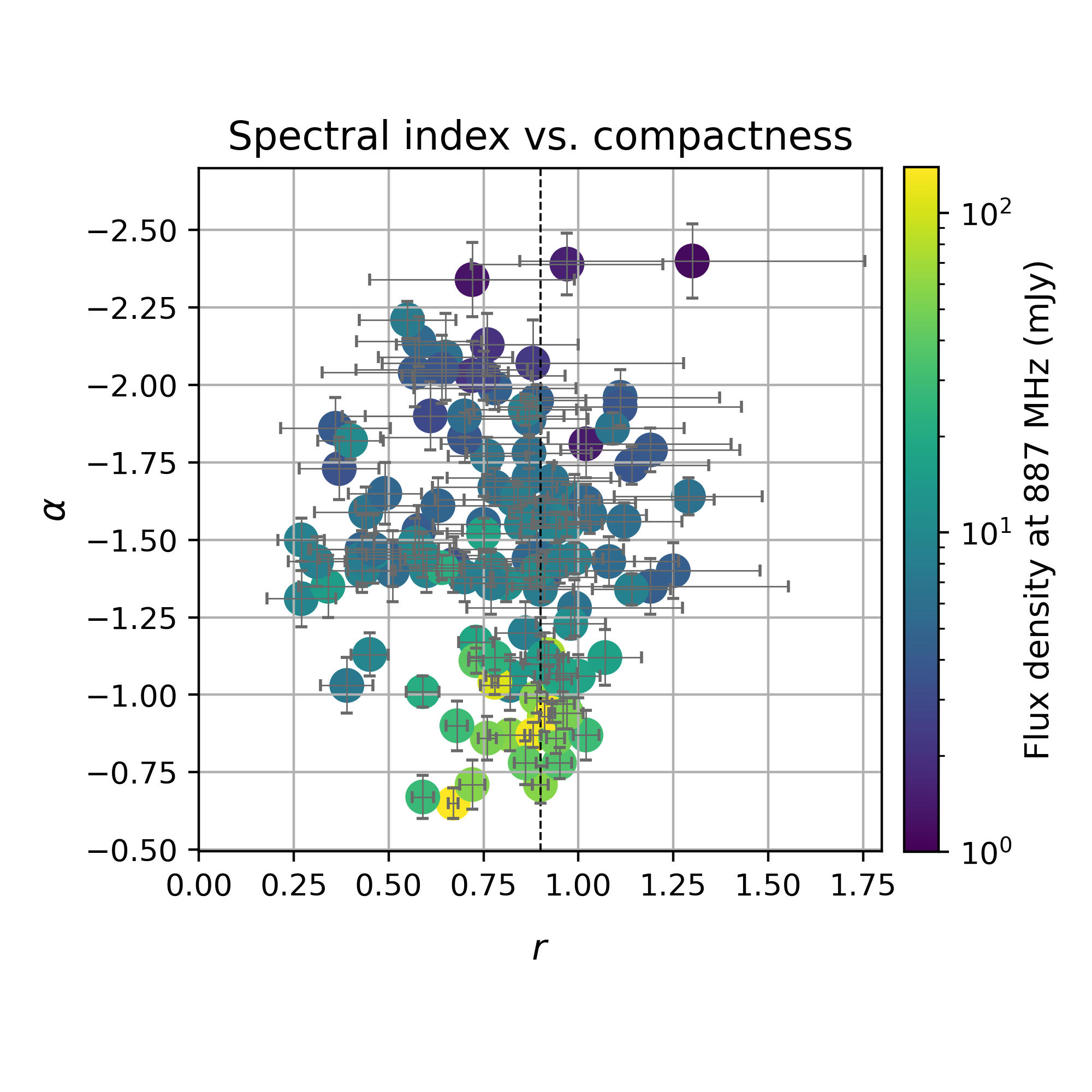}
        \caption{A scatter illustration of the spectral index ($\alpha$) vs the compactness parameter (\textit{r}) for the 118 sources classified as \textit{compact} and \textit{non-compact}, with a vertical line at \textit{r}=0.9 separating the two. The colour map shows their flux densities as observed in RACS-low. All sources with $r>1$ are consistent with $r=1$ within $1.5\sigma$ uncertainty. }
        \label{fig: scatterplot}
\end{figure}

In the preceding sections, we have presented spectral and compactness characterisation of 171 supposedly steep spectrum sources. For compactness, we have used qualitative as well as quantitative criteria. For example, resolved sources are identified visually --- if a source appears to be composed of multiple components or has a size clearly much larger than the synthesised beam, it is classified as \textit{resolved}. On the other hand, the compactness parameter based classification resulted in the identification of the \textit{non-compact} sources. For some sources, as shown by an example in Figure \ref{fig:3}, the distinction between \textit{non-compact} and \textit{resolved} might not be trivial. But, effectively, majority of the \textit{non-compact} sources will likely appear as resolved in observations with higher angular resolutions. For the \textit{resolved} as well as \textit{non-compact} sources, we have reported a \emph{global} spectral index by using \textit{ifd} (for \textit{resolved} sources, after convolving the RACS images to 25" $\times$ 25" TGSS resolution). A detailed spectral index map obtained using higher angular resolution images at multiple frequencies would provide richer details. Furthermore, we probed two higher resolution wide-field surveys at 1.4\,GHz \citep[Faint Images of the Radio Sky at Twenty-cm (FIRST);][]{FIRST} and 3\,GHz \citep[Very Large Array Sky Survey (VLASS);][]{VLASS} to inspect the 66 compact sources. A search through the FIRST catalogue revealed detections of only two sources, namely 094019+364153 and 164347-342413. Both remain compact, with sizes well below 10". A very faint emission is observed in VLASS at the location of 094019+364153, but not at a significance level to comment more on its morphological features at the survey's 3" resolution.

\begin{figure*}
    \centering

    \begin{minipage}[b]{\textwidth}
        \centering
        \textbf{Column A:} 17:50:03 $-$ 27:48:16 \hfill
        \textbf{Column B:} 18:09:43 $-$ 19:49:13 \hfill
        \textbf{Column C:} 18:27:48 $-$ 13:01:53
    \end{minipage}

    \vspace{0.75em}

    \includegraphics[width=0.331\textwidth]{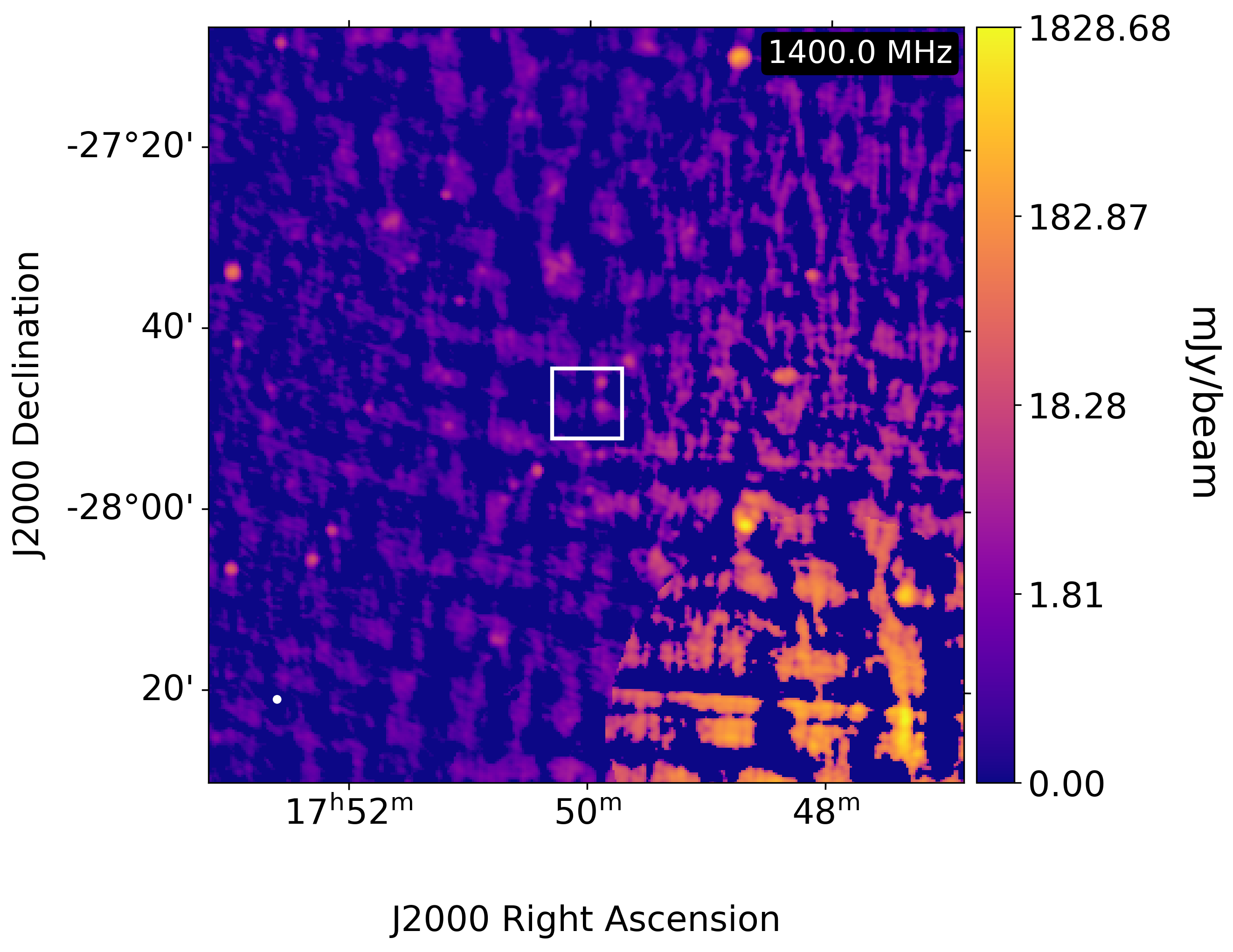} \hfill
    \includegraphics[width=0.331\textwidth]{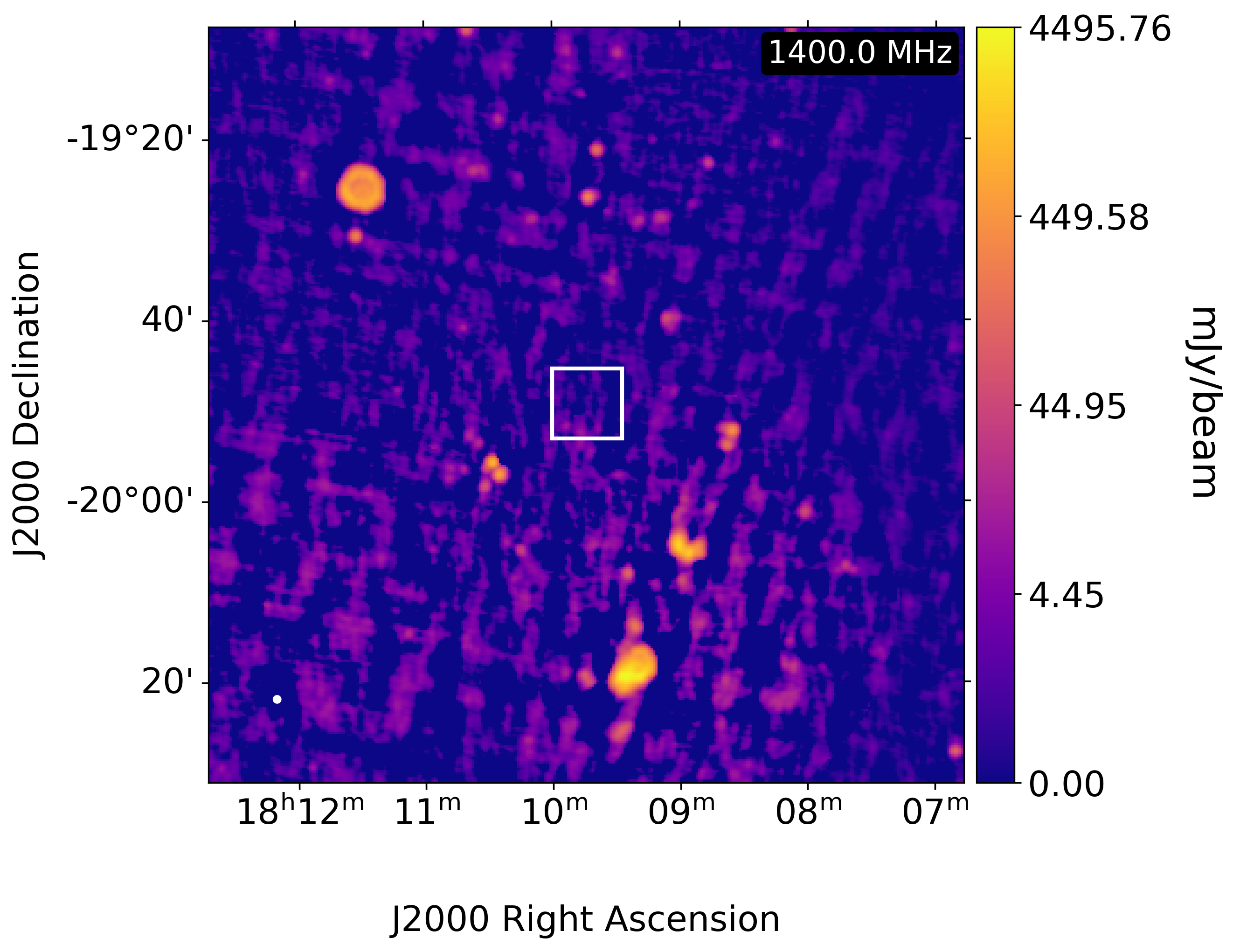} \hfill
    \includegraphics[width=0.331\textwidth]{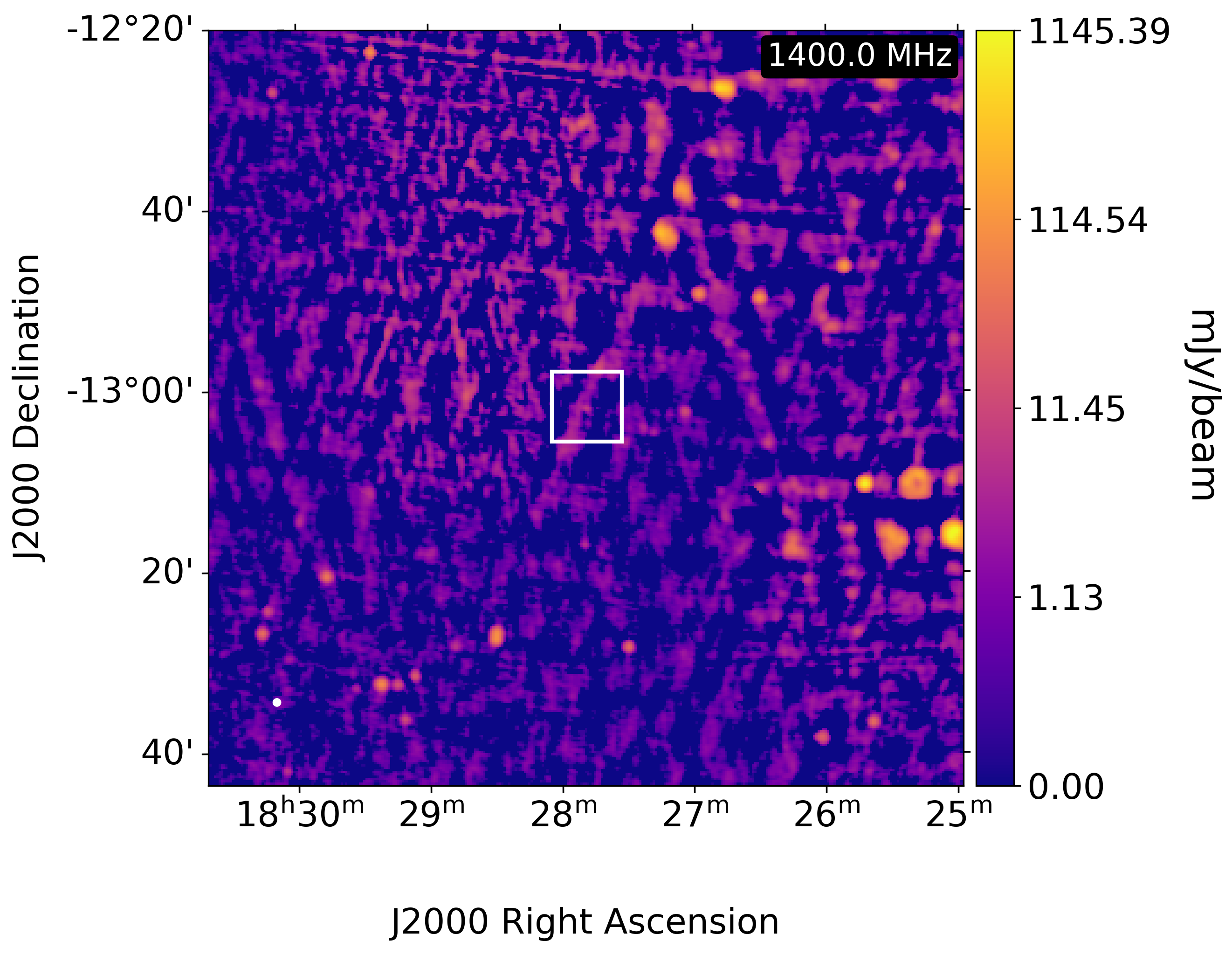}

    \par\vspace{0.5em}
    
    \textbf{(a) NVSS in the Galactic plane:} Wide field images of the Galactic plane centred at the selected coordinates. The images show severe contamination by imaging artefacts.

    \vspace{0.75em}
    \par\vspace{0.75em}

    \includegraphics[width=0.331\textwidth]{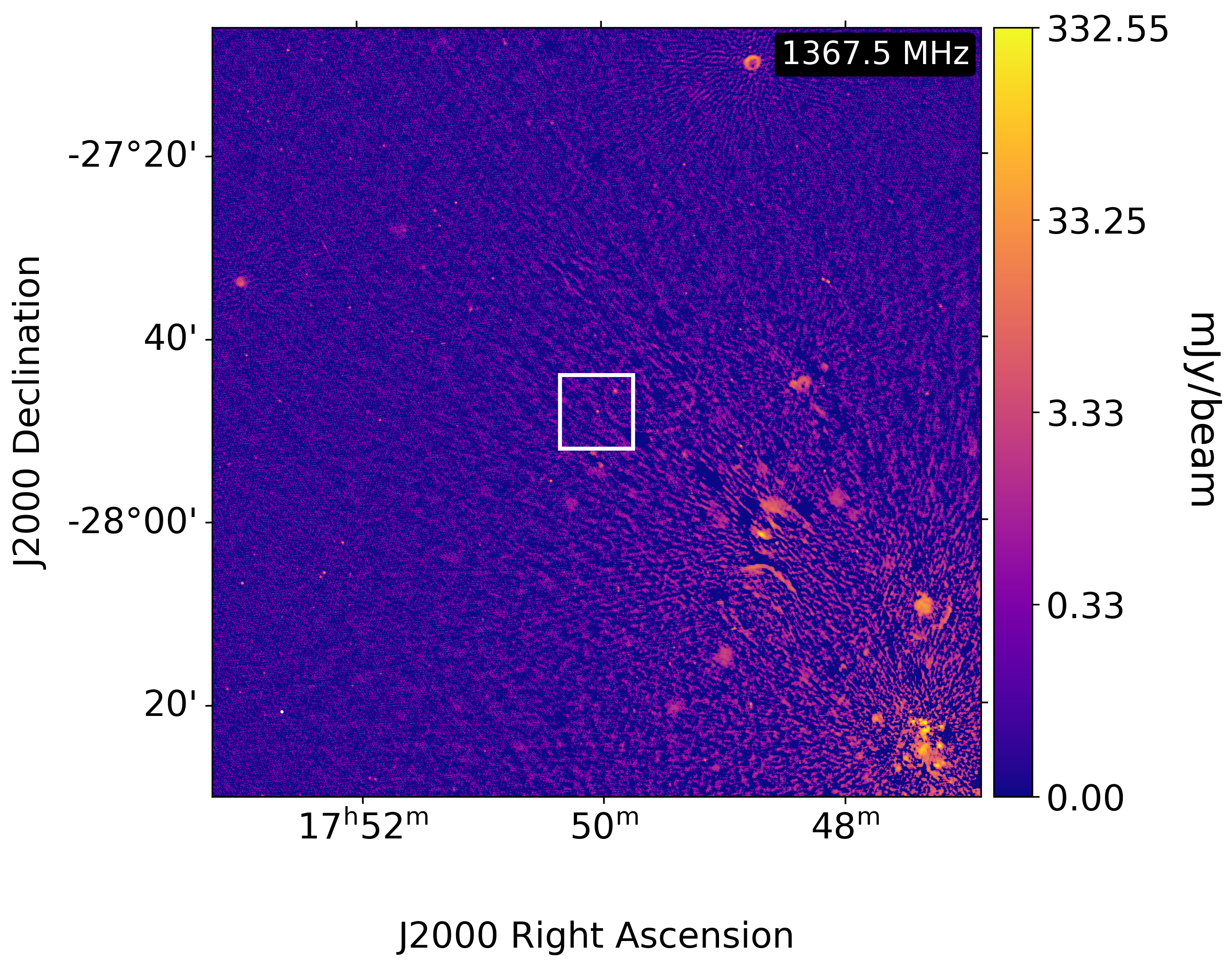} \hfill
    \includegraphics[width=0.331\textwidth]{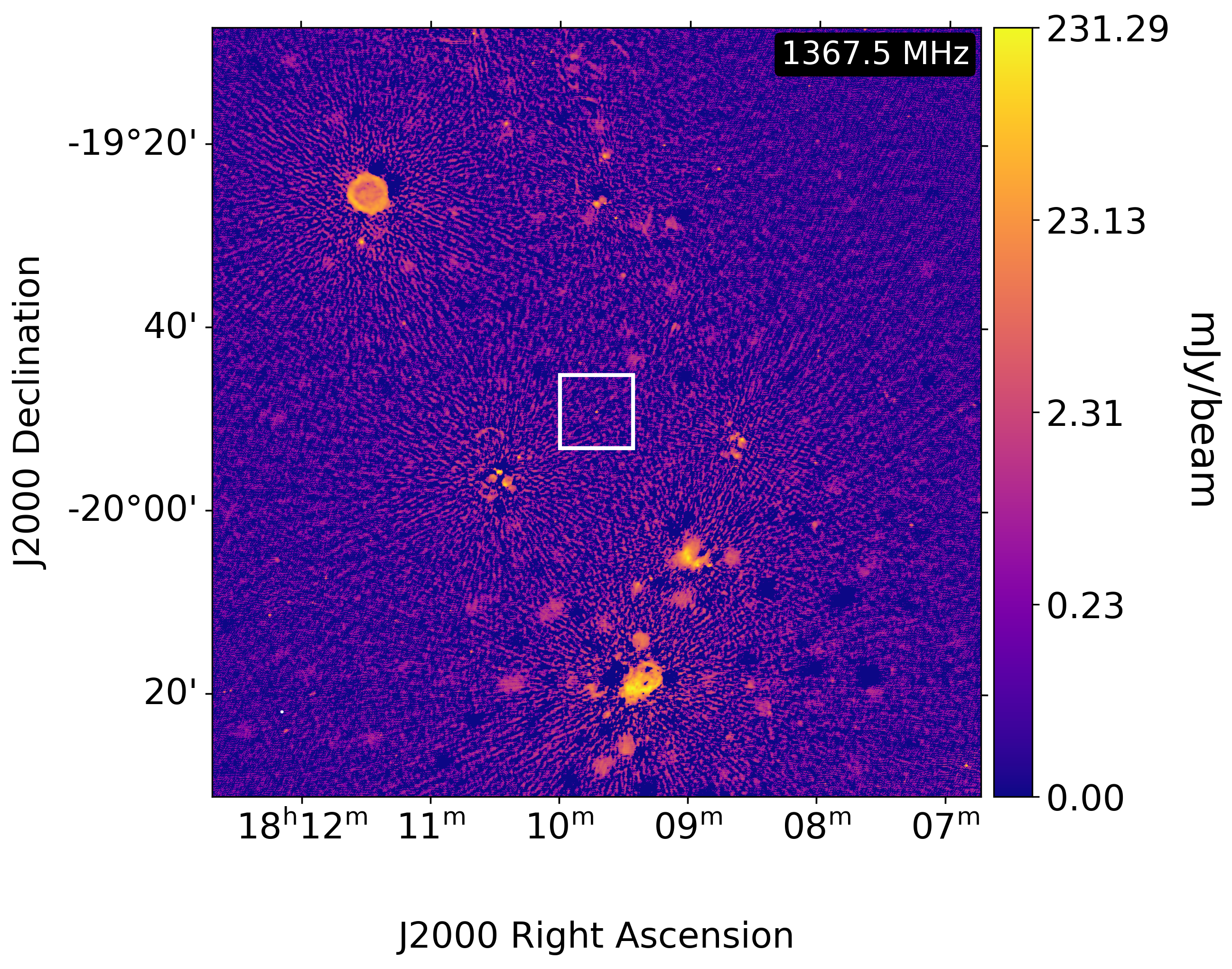} \hfill
    \includegraphics[width=0.331\textwidth]{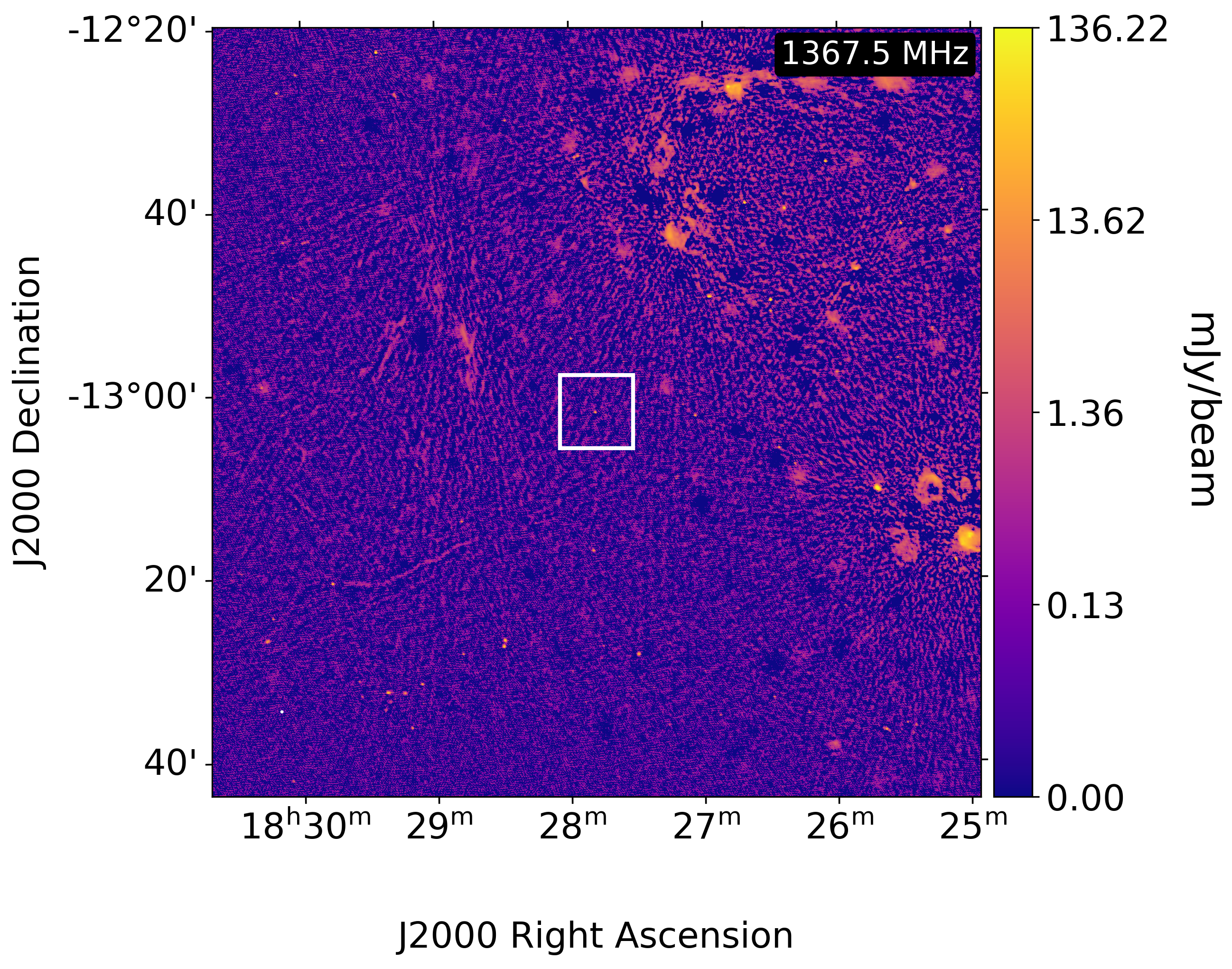}

    \par\vspace{0.3em}


    \includegraphics[width=0.331\textwidth]{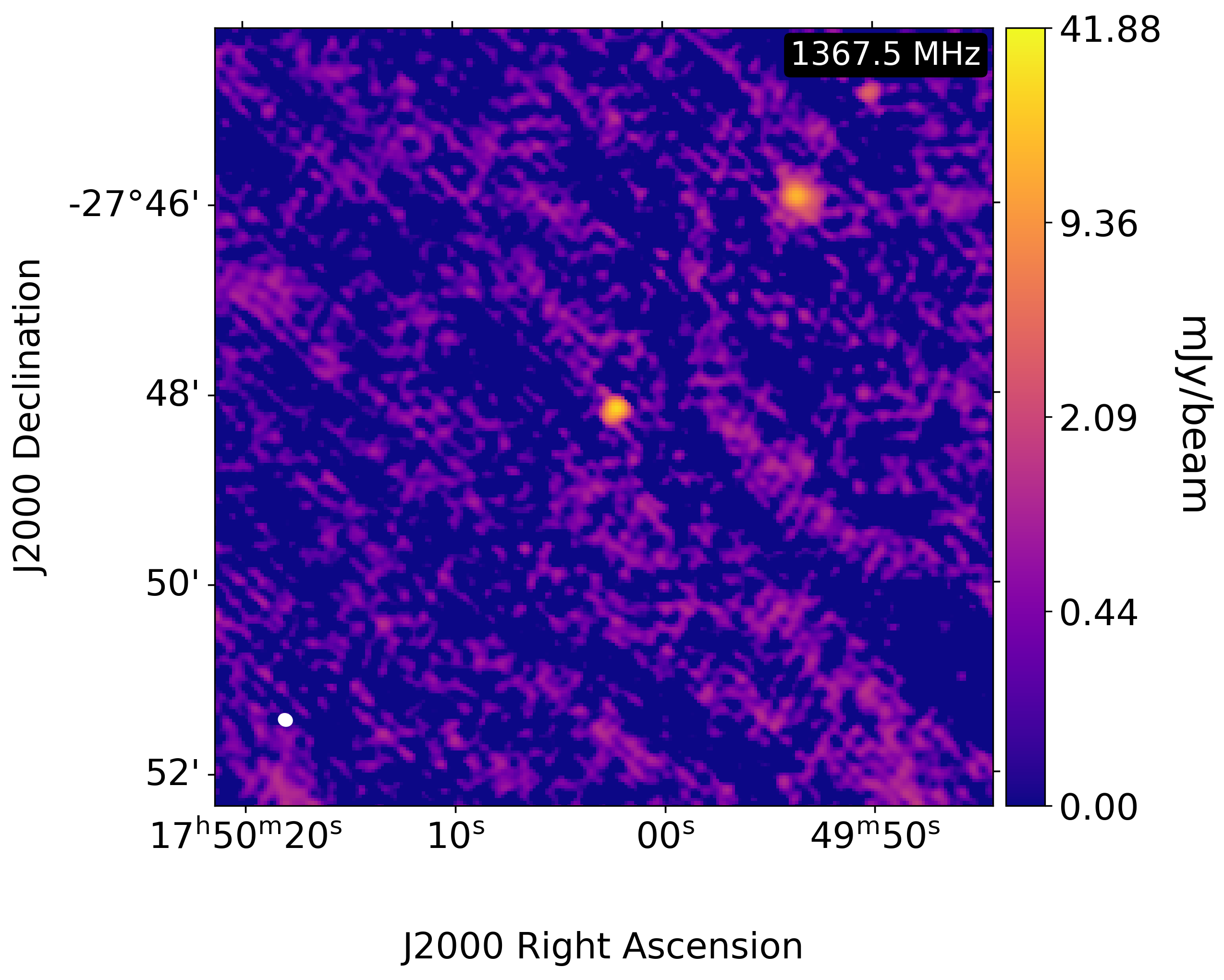} \hfill
    \includegraphics[width=0.331\textwidth]{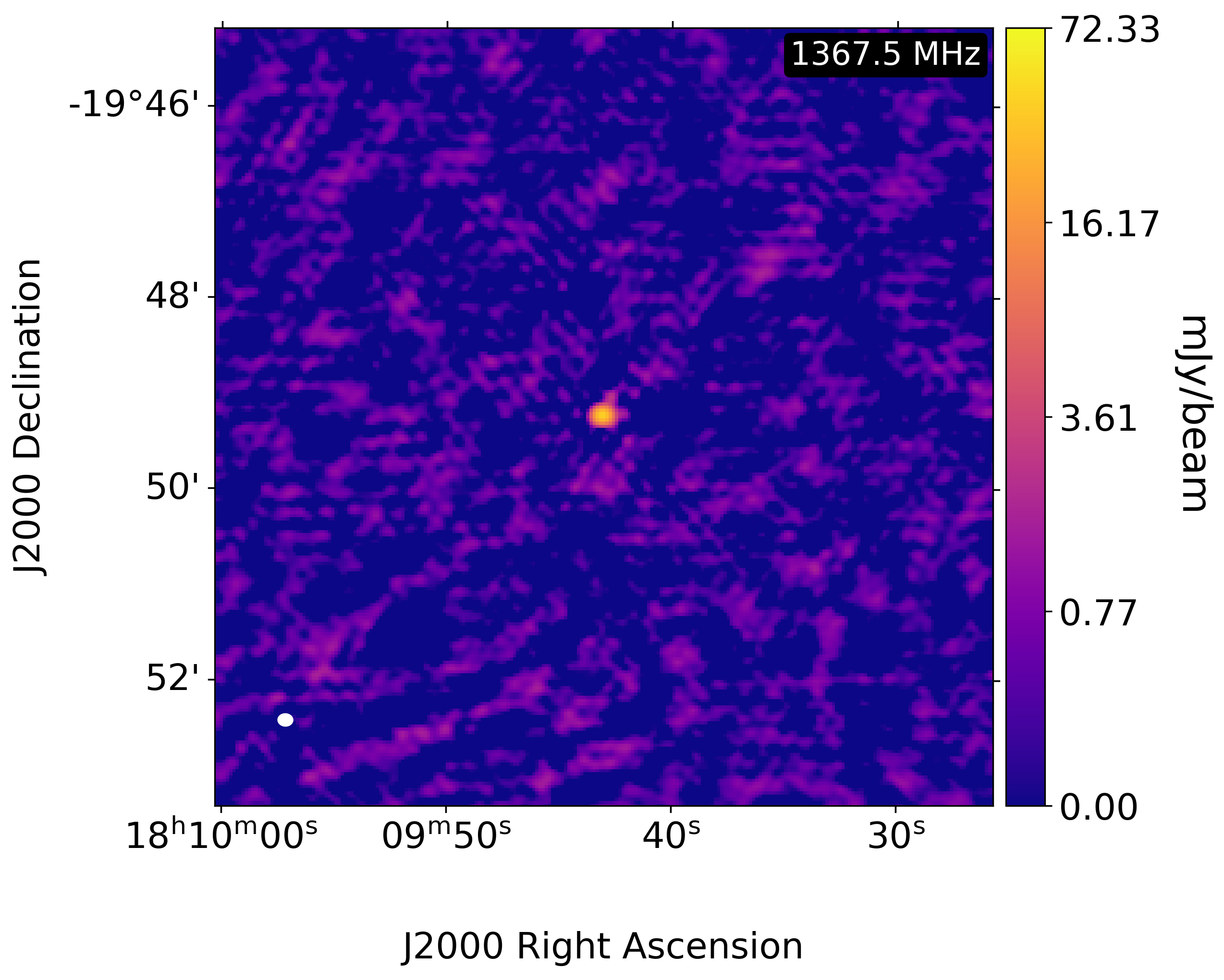} \hfill
    \includegraphics[width=0.331\textwidth]{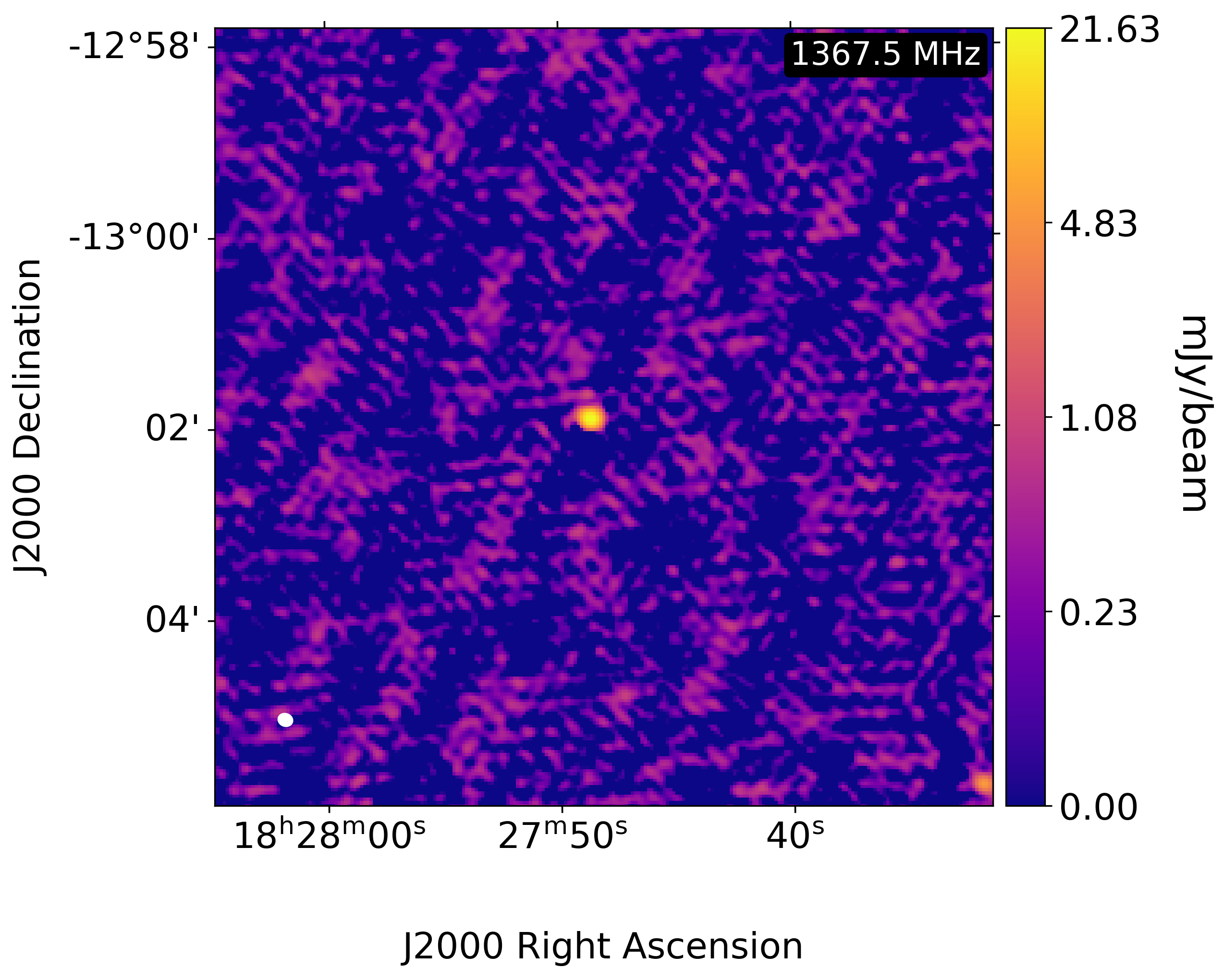}

    \par\vspace{0.75em}
    \textbf{(b) RACS Mid-Band (1367 MHz):} Higher resolution wide field (upper row) and zoomed-in (lower row) images centred at the same coordinates as in (a), with the bottom row showing clearly visible bright and compact sources. 

    \caption{Multi-resolution view of the Galactic plane at three different coordinates (columns A, B and C), illustrating image quality improvement from NVSS (top row; a) to RACS-mid (middle row; b). The bottom row (b) presents zoomed-in views of the middle-row images, where the sources are clearly visible. These zoomed regions correspond to the areas marked by white boxes in the wide-field images in the middle and upper rows. The sky coordinates of the sources at which these images are centred at, are shown above each of the individual columns.}
    \label{fig: NVSS vs RACS}
\end{figure*}

In Figure \ref{fig: scatterplot}, we present the estimated spectral indices against the compactness parameter, \textit{r}, along with the flux densities indicated by colour, for all the sources classified as \textit{compact} and \textit{non-compact}. We chose the flux densities at 887\,MHz as some of the sources are not detected at 1367\,MHz. While most of these spectral indices are from fits using three data points, some of them are two-point spectral indices (as indicated in Tables \ref{tab:table1} and \ref{tab:table2}; where the flux density measurements were available only at two frequencies). For compact sources with flux density measurements at three frequencies, we did not find any measurable evidence for a curvature in the spectrum. We assume that spectra of the remaining compact sources are also well described by power laws. In case of extended sources our estimates of spectral indices are not robust as different baseline coverage for different measurements could lead to missing flux at certain frequencies.

We observe a a trend between the spectral index and the flux density. In general, the brighter the source is, the flatter the spectral index --- it is not entirely surprising though, as this trend would be expected unless the TGSS flux density distribution of the sources in our sample is very broad. What is more surprising is that the spectral indices for a significant number of sources is flatter than -1.5, the limit with which these sources were selected in the first place. As the previous spectral index limits came from the corresponding NVSS flux density limits, the sources with significantly flatter spectral indices might indicate towards limitations of NVSS in those directions.
 
\par
To further probe the origin of the flatter spectral indices, we took a careful look at some of the sources that exhibited discrepant results.
Figure ~\ref{fig: NVSS vs RACS}(a) shows NVSS images centred at three sources, 175003-274816, 180943-194913 and 182748-130153, with Figure ~\ref{fig: NVSS vs RACS}(b) showing the corresponding RACS-mid images. All the three sources lie in the Galactic plane. In addition to the large field of view images, Figure ~\ref{fig: NVSS vs RACS}(b) also shows zoomed-in views which clearly reveal the individual sources, with peak flux densities of 26.6\,mJy\,beam$^{-1}$, 42.6\,mJy\,beam$^{-1}$ and 26.5\,mJy\,beam$^{-1}$ at 1367\,MHz (RACS-mid). However, these sources are not visible at 1400\,MHz in the NVSS images, owing to the strong and broad artefacts in their surroundings. It is likely that the local rms estimation in such cases might not have appropriately taken into account the broad and widespread artefacts, and hence, provides rms values much smaller than the actual representative values. The above three sources had spectral index upper limits of -1.52, -1.91 and -1.58, respectively, which are significantly steeper than what we have found in this work (-0.71$\pm$0.08, -1.13$\pm$0.04 and -0.86$\pm$0.05, respectively). We identified several other similar examples, such as the sources 183113-021012, 203401+401009, 203020+382337, etc., in our GP sample. While the RACS images also suffer from systematics in the Galactic plane, with the higher angular resolution, it is much easier to take those into account while estimating the local rms. The poor image quality in NVSS, particularly in the Galactic plane, thus requires a more careful noise analysis before the flux densities or limits therefrom could be reliably used for any scientific purpose.

\begin{figure*}
    \centering
    \includegraphics[width=1.0\textwidth]{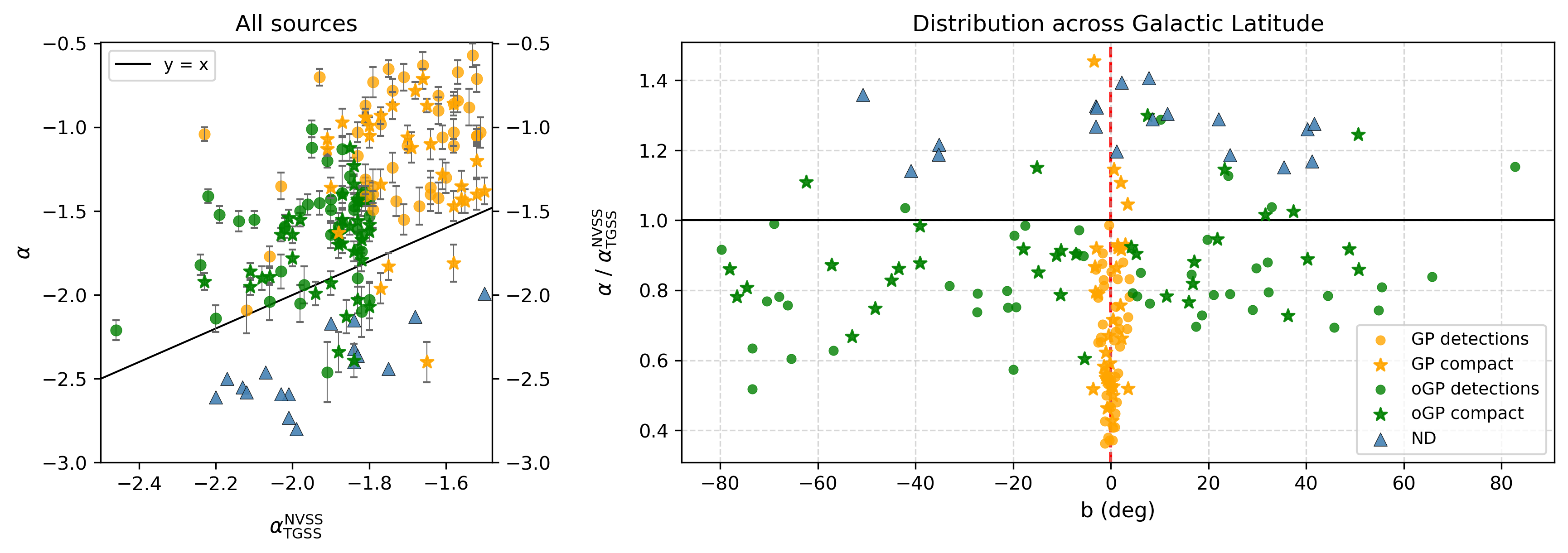}
    \caption{\textit{Left:} Scatter diagram showing the upper limits on the spectral indices from \citet[][using TGSS--NVSS]{deGasperin18} vs our estimates using TGSS and RACS, for all the 171 sources in our sample. The orange and green points indicate the sources in the GP and oGP samples, respectively, which were detected in TGSS, RACS-low as well as RACS-mid. Source which were not detected in RACS (\textit{ND}) are shown as blue triangles. \textit{Right:} The ratio of our spectral indices 
    to the limits from \citet{deGasperin18} are shown for all the source as a function of their Galactic latitudes. The dashed red line indicates the zero Galactic latitude, i.e., $b$=0 deg.}
    \label{fig: alpha_vs_alpha}
\end{figure*}

The imaging artefacts are more pronounced in the Galactic plane due to the presence of the large number density of brighter sources. However, even outside the galactic plane, we note that TGSS$-$NVSS provided spectral index limits which are typically 15$-$20\% steeper than the actual indices estimated using the TGSS$-$RACS pair (see Figure ~\ref{fig: alpha_vs_alpha}). The reasons for this discrepancy could be multi-fold. In many cases, a visual examination of the NVSS images reveals faint emission at the corresponding positions, suggestive of the sources being real but at a significance level that was insufficient for inclusion in the \citet{deGasperin18} catalogue (i.e., $<4\sigma_l$). With the better sensitivity, RACS-mid clearly detects such sources and provides the spectral index estimates that are slightly flatter than their TGSS$-$NVSS limits. In some other cases, broad artefacts affect the estimates even outside the Galactic plane, although the impact is relatively smaller. A visual inspection of sources in the surveys like NVSS is thus important to assess such effects, an aspect which was also highlighted by \cite{Frail2017ImageSearch}. Additionally, we note that such a discrepancy in spectral index estimates could also arise from a spectral curvature in some cases, where estimates using different pairs of frequencies would not agree with each other. However, for compact sources with flux density measurements at three frequencies, we did not find any measurable evidence for a curvature in the spectrum. It is hence likely that spectra of the majority of the remaining compact sources would also be well described by power-laws. 

\par
Variability between the NVSS and RACS observation epochs might also produce inconsistencies in the derived spectral limits/indices. Strong variables are believed to constitute around $1\%$ of the radio sources at 1.4\,GHz \citep{Hancock2016}, and a much lesser fraction at 150\,MHz \citep{Hajela2019}. Such variability can impact the spectral indices typically by $\pm 0.15$ \citep{Frail2017ImageSearch}. However, we suspect this effect to be secondary and might have affected only a small fraction of our sources. 

\par
For all the sources which were not detected, i.e., classified as \textit{ND}, we obtain spectral index limits steeper than the limits provided by \citet{deGasperin18} using TGSS$-$NVSS (see Figure ~\ref{fig: alpha_vs_alpha}). This is only natural, given the typically better sensitivity of RACS when compared to NVSS. Figure ~\ref{fig: alpha_vs_alpha} also highlights that majority of the steepest spectral indices produced in our work are accounted for by \textit{ND}s. Short-lived transients or variable sources can potentially explain some of these sources, resulting in anomalously steep indices, although the fraction of such sources is expected to be very small. Furthermore, up to $\sim 5\%$ of the upper-limits in \cite{deGasperin18} are expected to be represented by false detections associated with imaging artefacts. These could account for some of the \textit{ND}s identified in our sample. However, a visual examination of all the \textit{ND} source positions in TGSS reveals that majority of these sources are indeed real and unlikely to be the result of artefacts. These sources could be likely pulsars as many known pulsars exhibit similarly steep or even steeper spectra (intrinsically or due to geometric effects).

\par
Finally, our characterisation of the spectral indices and compactness of all the sources (see Appendix) following close inspection of the individual RACS images provide refined and useful starting points for a variety of studies. It might be interesting to probe if some of the clearly resolved and multi-component sources are intriguing active galactic nuclei or luminous radio galaxies. The compact sources with steep or ultra-steep spectra remain strong candidates for new pulsars and HzRGs \citep{Maan18}. Just on the basis of the spectral index limits and compactness obtained from TGSS and NVSS, all the 171 sources in our original sample were pulsar candidates. We have shown that higher angular resolution observations, in this case available from RACS, can perform an important role in refining such initial samples --- only 56 sources which are either found to be compact or not-detected in RACS, and have $\alpha<-1.5$, remain promising pulsar candidates (38 compact and 18 ND). In fact, two sources (190134-012527 and 050922+085625) were already known to be pulsars at the time of this work but remained included in our sample due to usage of an older version of the \texttt{psrcat}. Both of these sources have been identified as compact and steep or ultra-steep spectrum sources in this work. Also, one of the sources in our sample corresponds to the M28 globular cluster which is already known to host several pulsars but the source was still kept in our sample. The source is identified with an ultra steep spectrum ($\alpha=-2.21\pm0.06$), but, interestingly, it appears as \textit{non-compact} ($r=0.55\pm0.13$). The non-compactness in this particular case might have originated from multiple radio pulsars and other sources (e.g., stellar mass black holes) within the globular cluster, contributing to the total integrated flux density. Furthermore, in a survey of compact sources for pulsars and exotic objects \citep[SCOPE\footnote{SCOPE webpage: \url{http://www.ncra.tifr.res.in/~ymaan/scope.html}};][]{Maan26a}, pulsation searches of a subset of compact sources with steep spectra identified from this work have already resulted in the discovery of two pulsars towards the sources 184009+110207 and 182736-044941. Both sources have ultra-steep spectra (with $\alpha$ of -2.55 and -2.39). The discovery of these two pulsars further demonstrate the usefulness of this work and motivate pulsation searches of all the compact candidates with steep spectra revealed here. It would also be useful to probe these sources with higher angular resolutions and at optical wavelengths to uncover any underlying HzRGs.
\section{Conclusions}\label{Conclusion}

Using the publicly available RACS images, we have presented a detailed characterisation of an all-sky sample comprising of 171 sources, identified as compact radio sources with steep-spectra using information from TGSS and NVSS. The characterisation involved estimating the spectral indices using the flux density measurements at as many frequencies as possible, as well as the compactness from the highest frequency detection of the individual sources. The estimated spectral indices, compactness as well as a morphological classification that could be useful for physical identification of various categories of the sources, are presented for the whole sample. This work demonstrates that, especially in the Galactic plane, a more careful noise analysis of the NVSS data products is needed before they can be used to identify truly compact sources with steep spectra. Our usage of data products from a higher angular resolution and sensitive survey, RACS, provided a more accurate estimation of spectral indices and a far better identification of compact sources with steep spectra. The complete characterisation of all sources, is made available in the form of tables. Overall, the characterisation presented in this work provides useful starting points to probe different categories of exciting sources, including pulsars as well as HzRGs.

\begin{acknowledgement}
This scientific work uses data obtained from Inyarrimanha Ilgari Bundara / the Murchison Radio-astronomy Observatory. We acknowledge the Wajarri Yamaji People as the Traditional Owners and native title holders of the Observatory site. CSIRO’s ASKAP radio telescope is part of the Australia Telescope National Facility (https://ror.org/05qajvd42). Operation of ASKAP is funded by the Australian Government with support from the National Collaborative Research Infrastructure Strategy. ASKAP uses the resources of the Pawsey Supercomputing Research Centre. Establishment of ASKAP, Inyarrimanha Ilgari Bundara, the CSIRO Murchison Radio-astronomy Observatory and the Pawsey Supercomputing Research Centre are initiatives of the Australian Government, with support from the Government of Western Australia and the Science and Industry Endowment Fund. This paper includes archived data obtained through the CSIRO ASKAP Science Data Archive, CASDA (https://data.csiro.au). We thank the GMRT staff that made these observations possible. GMRT is run by the National Centre for Radio Astrophysics of the Tata Institute of Fundamental Research. We also use data from the NRAO VLA Sky Survey \citep{Condon98}. 
\end{acknowledgement}

\paragraph{Funding Statement}
RS acknowledges support from the Kishore Vaigyanik Protsahan Yojana (KVPY) fellowship. YM acknowledges support from the Department of Science and Technology via the Science and Engineering Research Board Startup Research Grant (SRG/2023/002657), as well as the Department of Atomic Energy for funding support, under project 12$-$R\&D$-$TFR$-$5.02$-$0700. AB acknowledges support through project CORTEX (NWA.1160.18.316) of the research programme NWA-ORC which is financed by the Dutch Research Council (NWO).

\paragraph{Competing Interests}
None.
\paragraph{Data Availability Statement}
Data will be made available on request.

\clearpage
\bibliography{ref}

@ARTICLE{BLV13,
   author = {{Bates}, S.~D. and {Lorimer}, D.~R. and {Verbiest}, J.~P.~W.},
    title = "{The pulsar spectral index distribution}",
  journal = "MNRAS",
     year = 2013,
   volume = 431,
    pages = {1352-1358},
}

@ARTICLE{Maan26a,
       author = {{Maan}, Yogesh and {Bera}, Apurba and {Vir Lal}, Dharam and {Bhusare}, Yash and {Kharb}, Preeti and {Lal}, Banshi and {Atri}, Pikky},
        title = "{Survey of compact sources for pulsars and exotic objects -- I. Overview and initial discoveries}",
      journal = {arXiv e-prints},
         year = 2026,
        month = mar,
          eid = {arXiv:2603.28885},
        pages = {arXiv:2603.28885},
          doi = {10.48550/arXiv.2603.28885},
archivePrefix = {arXiv},
       eprint = {2603.28885},
 primaryClass = {astro-ph.HE},
       adsurl = {https://ui.adsabs.harvard.edu/abs/2026arXiv260328885M}
}

@article{Frail2017ImageSearch,
  author  = {Frail, D. A. and Ray, P. S. and Mooley, K. P. and Hancock, P. and Burnett, T. H. and Jagannathan, P. and Ferrara, E. C. and Intema, H. T. and de Gasperin, F. and Demorest, P. B. and Stovall, K. and McKinnon, M. M.},
  title   = {An image‑based search for pulsars among Fermi unassociated LAT sources},
  journal = {MNRAS},
  volume  = {475},
  pages   = {942--954},
  year    = {2017},
  doi     = {10.1093/mnras/stx3281}
}

@ARTICLE{Navarro95,
   author = {{Navarro}, J. and {de Bruyn}, A.~G. and {Frail}, D.~A. and {Kulkarni}, S.~R. and {Lyne}, A.~G.},
    title = "{A Very Luminous Binary Millisecond Pulsar}",
  journal = "ApJL",
     year = 1995,
    month = dec,
   volume = 455,
    pages = {L55},
      doi = {10.1086/309816},
}

@article{Kaplan1998NVSS,
  author  = {Kaplan, David L. and Condon, James J. and Arzoumanian, Zaven and Cordes, James M.},
  title   = {Pulsars in the NRAO VLA Sky Survey},
  journal = {ApJS},
  volume  = {119},
  pages   = {75--??},
  year    = {1998},
  doi     = {10.1086/313153}
}

@article{McConnell2020,
  author = {McConnell, D. and Hale, C. L. and Lenc, E. and Murphy, T.},
  title = {The Rapid ASKAP Continuum Survey I: Design and First Results},
  journal = {PASA},
  volume = {37},
  pages = {e048},
  year = {2020},
  doi = {10.1017/pasa.2020.41}
}

@ARTICLE{atnf,
   author = {{Manchester}, R.~N. and {Hobbs}, G.~B. and {Teoh}, A. and {Hobbs}, M.},
    title = "{The Australia Telescope National Facility Pulsar Catalogue}",
  journal = "AJ",
     year = 2005,
   volume = 129,
    pages = {1993-2006},
}

@ARTICLE{DeBreuck00,
   author = {{De Breuck}, C. and {van Breugel}, W. and {R{\"o}ttgering}, H.~J.~A. and {Miley}, G.},
    title = "{A sample of 669 ultra steep spectrum radio sources to find high redshift radio galaxies}",
  journal = {A\&AS},
   eprint = {astro-ph/0002297},
     year = 2000,
    month = apr,
   volume = 143,
    pages = {303-333},
      doi = {10.1051/aas:2000181},
   adsurl = {http://adsabs.harvard.edu/abs/2000A%26AS..143..303D}
}

@ARTICLE{Intema17,
   author = {{Intema}, H.~T. and {Jagannathan}, P. and {Mooley}, K.~P. and {Frail}, D.~A.},
    title = "{The GMRT 150 MHz all-sky radio survey. First alternative data release TGSS ADR1}",
  journal = "A\&A",
     year = 2017,
    month = feb,
   volume = 598,
      eid = {A78},
    pages = {A78},
      doi = {10.1051/0004-6361/201628536},
}

@ARTICLE{Frail16,
   author = {{Frail}, D.~A. and {Jagannathan}, P. and {Mooley}, K.~P. and {Intema}, H.~T.},
    title = "{Known Pulsars Identified in the GMRT 150 MHz All-sky Survey}",
  journal = "ApJ",
     year = 2016,
   volume = 829,
    pages = {119},
}

@ARTICLE{Condon98,
   author = {{Condon}, J.~J. and {Cotton}, W.~D. and {Greisen}, E.~W. and {Yin}, Q.~F. and {Perley}, R.~A. and {Taylor}, G.~B. and {Broderick}, J.~J.},
    title = "{The NRAO VLA Sky Survey}",
  journal = "AJ",
     year = 1998,
   volume = 115,
    pages = {1693-1716},
      doi = {10.1086/300337},
}

@article{Han1999SNR,
  author  = {Han, J. L. and Tian, W. W.},
  title   = {An updated catalog of pulsars with associations to supernova remnants},
  journal = {ApJS},
  volume  = {136},
  pages   = {571--582},
  year    = {1999}
}

@ARTICLE{Backer82,
   author = {{Backer}, D.~C. and {Kulkarni}, S.~R. and {Heiles}, C. and {Davis}, M.~M. and {Goss}, W.~M.},
    title = "{A millisecond pulsar}",
  journal = "Nature",
     year = 1982,
    month = dec,
   volume = 300,
    pages = {615-618},
      doi = {10.1038/300615a0},
}

@ARTICLE{Lyne87,
   author = {{Lyne}, A.~G. and {Brinklow}, A. and {Middleditch}, J. and {Kulkarni}, S.~R. and {Backer}, D.~C.},
    title = "{The discovery of a millisecond pulsar in the globular cluster M28}",
  journal = "Nature",
     year = 1987,
    month = jul,
   volume = 328,
    pages = {399-401},
      doi = {10.1038/328399a0},
}

@ARTICLE{Hamilton85,
   author = {{Hamilton}, T.~T. and {Helfand}, D.~J. and {Becker}, R.~H.},
    title = "{A search for millisecond pulsars in globular clusters}",
  journal = "AJ",
     year = 1985,
    month = apr,
   volume = 90,
    pages = {606-608},
      doi = {10.1086/113767},
   adsurl = {http://adsabs.harvard.edu/abs/1985AJ.....90..606H},
}

@ARTICLE{Marthi11,
   author = {{Marthi}, V.~R. and {Chengalur}, J.~N. and {Gupta}, Y. and {Dewangan}, G.~C. and {Bhattacharya}, D.},
    title = "{The central point source in G76.9+1.0}",
  journal = "MNRAS",
     year = 2011,
    month = oct,
   volume = 416,
    pages = {2560-2566},
      doi = {10.1111/j.1365-2966.2011.19155.x},
}

@ARTICLE{Bhakta17,
   author = {{Bhakta}, D. and {Deneva}, J.~S. and {Frail}, D.~A. and {de Gasperin}, F. and {Intema}, H.~T. and {Jagannathan}, P. and {Mooley}, K.~P.},
    title = "{Searching for pulsars associated with the Fermi GeV excess}",
  journal = "MNRAS",
     year = 2017,
    month = jul,
   volume = 468,
    pages = {2526-2531},
      doi = {10.1093/mnras/stx656},
}

@ARTICLE{Maan18,
       author = {{Maan}, Yogesh and {Bassa}, Cees and {van Leeuwen}, Joeri and
         {Krishnakumar}, M.~A. and {Joshi}, Bhal Chandra},
        title = "{A Search for Pulsars in Steep Spectrum Radio Sources}",
      journal = "ApJ",
         year = "2018",
        month = "Sep",
       volume = {864},
       number = {1},
          eid = {16},
        pages = {16},
          doi = {10.3847/1538-4357/aad4ad},
archivePrefix = {arXiv},
       eprint = {1807.08363},
 primaryClass = {astro-ph.HE},
       adsurl = {https://ui.adsabs.harvard.edu/abs/2018ApJ...864...16M}
}

@ARTICLE{Tiwari19,
  author = {{Tiwari}, Prabhakar},
  title = "{Radio spectral index from NVSS and TGSS}",
  journal = {RAA},
  year = {2019},
  volume = {19},
  number = {7},
  pages = {96},
  doi = {10.1088/1674-4527/19/7/96},
  eprint = {1609.01308},
  archivePrefix = {arXiv},
  primaryClass = {astro-ph.GA}
}

@ARTICLE{deGasperin18,
   author = {{de Gasperin}, F. and {Intema}, H.~T. and {Frail}, D.~A.},
    title = "{A radio spectral index map and catalogue at 147-1400 MHz covering 80 per cent of the sky}",
  journal = "MNRAS",
     year = 2018,
    month = mar,
   volume = 474,
    pages = {5008-5022},
      doi = {10.1093/mnras/stx3125},
}

@article{Hyman2021,
  author  = {Hyman, S. D. and Frail, D. A. and Deneva, J. S. and Kassim, N. E. and Giacintucci, S. and Kooi, J. E. and Lazio, T. J. W. and Joyner, I. and Peters, W. M. and Gajjar, V. and Siemion, A. P. V.},
  title   = {Two extreme steep-spectrum, polarized radio sources towards the Galactic bulge},
  journal = {MNRAS},
  volume  = {507},
  number  = {3},
  pages   = {3888--3898},
  year    = {2021},
  doi     = {10.1093/mnras/stab1979}
}

@ARTICLE{Damico85,
   author = {{Damico}, N. and {Manchester}, R.~N. and {Durdin}, J.~M. and {Erickson}, W.~C.},
    title = "{A search of steep-spectrum radio sources for millisecond pulsars}",
  journal = "{PASA}",
     year = 1985,
   volume = 6,
    pages = {174-176},
      doi = {10.1017/S1323358000018026},
}

@ARTICLE{Kaplan00,
   author = {{Kaplan}, D.~L. and {Cordes}, J.~M. and {Condon}, J.~J. and {Djorgovski}, S.~G.},
    title = "{Compact Radio Sources with the Steepest Spectra}",
  journal = "ApJ",
     year = 2000,
    month = feb,
   volume = 529,
    pages = {859-865},
      doi = {10.1086/308307},
}

@article{racs_paper2,
  author       = {Hale, C. L. and McConnell, D. and Andernach, H. and others},
  title        = {The Rapid ASKAP Continuum Survey Paper II: First Stokes I Source Catalogue Data Release},
  journal      = {PASA},
  year         = {2021},
  volume       = {38},
  pages        = {e058},
  doi          = {10.1017/pasa.2021.47}
}

@article{racs4,
  author  = {Duchesne, S. W. and Thomson, A. J. M. and Pritchard, J. and Lenc, E. and Moss, V. A. and McConnell, D. and Wieringa, M. H. and Whiting, M. T. and Wang, Z. and Wang, Y.},
  title   = {The Rapid ASKAP Continuum Survey IV: Continuum imaging at 1367.5 MHz and the first data release of RACS-mid},
  journal = {PASA},
  volume  = {40},
  pages   = {e034},
  year    = {2023},
  doi     = {10.1017/pasa.2023.31}
}

@article{racs5,
  author       = {Duchesne, S. W. and Grundy, J. A. and Heald, G. H. and Lenc, E. and Leung, J. K. and McConnell, D. and Murphy, T. and Pritchard, J. and Rose, K. and Thomson, A. J. M. and Wang, Y. and Wang, Z. and Whiting, M. T.},
  title        = {The Rapid ASKAP Continuum Survey V: Cataloguing the sky at 1367.5 MHz and the second data release of RACS-mid},
  journal      = {PASA},
  volume       = {41},
  pages        = {e003},
  year         = {2024},
  month        = {Jan},
  doi          = {10.1017/pasa.2023.60},
}

@article{Lourenco2024SMART,
  author    = {Lourenço, L. and Chippendale, A. P. and Indermuehle, B. and Moss, V. A. and Murphy, Tara and Galvin, T. J. and Hellbourg, G. and Hotan, A. W. and Lenc, E. and Whiting, M. T.},
  title     = {Survey and Monitoring of ASKAP’s RFI Environment and Trends I: Flagging Statistics},
  journal   = {PASA},
  year      = {2024},
  volume    = {41},
  pages     = {e012},
  doi       = {10.1017/pasa.2024.4}
}

@article{Murphy2013VAST,
  author    = {Murphy, Tara and Chatterjee, Shami and Kaplan, David L. and Banyer, Jay and Bell, Martin E. and Bignall, Hayley E. and Bower, Geoffrey C. and Cameron, Robert and Coward, David M. and Cordes, James M. and Croft, Steve and Curran, J. R. and Djorgovski, S. G. and Farrell, Sean A. and Frail, Dale A. and Gaensler, B. M. and Galloway, Duncan K. and Johnston, Simon and Kamble, Atish and Law, Casey J. and Lazio, J. W. and Lo, Kitty K. and Macquart, Jean‑Pierre and Rea, Nanda and Rebbapragada, Umaa and Reynolds, Cormac and Ryder, Stuart D. and Schmidt, Brian and Soria, Roberto and Stairs, Ingrid H. and Tingay, Steven J. and Torkelsson, Ulf and Wagstaff, Kiri and Walker, Mark and Wayth, Randall B. and Williams, Peter K. G.},
  title     = {VAST: An ASKAP Survey for Variables and Slow Transients},
  journal   = {PASA},
  year      = {2013},
  volume    = {30},
  pages     = {e006},
  doi       = {10.1017/pasa.2012.006}
}

@ARTICLE{casa,
  author = {{CASA Team} and Bean, Ben and Bhatnagar, Sanjay and Castro, Sandra and Donovan Meyer, Jennifer and Emonts, Bjorn and Garcia, Enrique and Garwood, Robert and Golap, Kumar and Gonzalez Villalba, Justo and Harris, Pamela and Hayashi, Yohei and Hoskins, Josh and Hsieh, Mingyu and Jagannathan, Preshanth and Kawasaki, Wataru and Keimpema, Aard and Kettenis, Mark and Lopez, Jorge and Marvil, Joshua and Masters, Joseph and McNichols, Andrew and Mehringer, David and Miel, Renaud and Moellenbrock, George and Montesino, Federico and Nakazato, Takeshi and Ott, Juergen and Petry, Dirk and Pokorny, Martin and Raba, Ryan and Rau, Urvashi and Schiebel, Darrell and Schweighart, Neal and Sekhar, Srikrishna and Shimada, Kazuhiko and Small, Des and Steeb, Jan-Willem and Sugimoto, Kanako and Suoranta, Ville and Tsutsumi, Takahiro and van Bemmel, Ilse M. and Verkouter, Marjolein and Wells, Akeem and Xiong, Wei and Szomoru, Arpad and Griffith, Morgan and Glendenning, Brian and Kern, Jeff},
  title = "{CASA, the Common Astronomy Software Applications for Radio Astronomy}",
  journal = {PASP},
  year = {2022},
  month = nov,
  volume = {134},
  number = {1041},
  eid = {114501},
  pages = {114501},
  doi = {10.1088/1538-3873/ac9642},
  archivePrefix = {arXiv},
  eprint = {2210.02276},
  primaryClass = {astro-ph.IM},
  adsurl = {https://ui.adsabs.harvard.edu/abs/2022PASP..134k4501C}
}

@INPROCEEDINGS{casda2017,
  author    = {Chapman, J.~M. and Dempsey, J. and Miller, D. and Heywood, I. and Pritchard, J. and Sangster, E. and Whiting, M. and Dart, M.},
  title     = {CASDA: The CSIRO ASKAP Science Data Archive},
  booktitle = {Astronomical Data Analysis Software and Systems XXV},
  year      = {2017},
  editor    = {Lorente, N.~P.~F. and Shortridge, K. and Wayth, R.},
  series    = {Astronomical Society of the Pacific Conference Series},
  volume    = {512},
  month     = dec,
  pages     = {73},
  adsurl    = {https://ui.adsabs.harvard.edu/abs/2017ASPC..512...73C}
}

@ARTICLE{Hancock2016,
       author = {{Hancock}, P.~J. and {Drury}, J.~A. and {Bell}, M.~E. and {Murphy}, T. and {Gaensler}, B.~M.},
        title = "{Radio variability in the Phoenix Deep Survey at 1.4 GHz}",
      journal = {MNRAS},
         year = 2016,
        month = sep,
       volume = {461},
       number = {3},
        pages = {3314-3321},
          doi = {10.1093/mnras/stw1486},
archivePrefix = {arXiv},
       eprint = {1606.05953},
 primaryClass = {astro-ph.GA},
       adsurl = {https://ui.adsabs.harvard.edu/abs/2016MNRAS.461.3314H}
}

@ARTICLE{Hajela2019,
       author = {{Hajela}, A. and {Mooley}, K.~P. and {Intema}, H.~T. and {Frail}, D.~A.},
        title = "{A GMRT 150 MHz search for variables and transients in Stripe 82}",
      journal = {MNRAS},
         year = 2019,
        month = dec,
       volume = {490},
       number = {4},
        pages = {4898-4906},
          doi = {10.1093/mnras/stz2918},
archivePrefix = {arXiv},
       eprint = {1910.10170},
 primaryClass = {astro-ph.IM},
       adsurl = {https://ui.adsabs.harvard.edu/abs/2019MNRAS.490.4898H}
}

@ARTICLE{ODeaSaikia21,
  author = {{O'Dea}, C.~P. and {Saikia}, D.~J.},
  title = "{Compact steep-spectrum and peaked-spectrum radio sources}",
  journal = {A\&ARv},
  year = {2021},
  volume = {29},
  number = {1},
  pages = {3},
  doi = {10.1007/s00159-021-00131-w},
  eprint = {2009.02750},
  archivePrefix = {arXiv},
  primaryClass = {astro-ph.GA}
}

@ARTICLE{HerzogCSS,
       author = {{Herzog}, A. and {Norris}, R.~P. and {Middelberg}, E. and {Seymour}, N. and {Spitler}, L.~R. and {Emonts}, B.~H.~C. and {Franzen}, T.~M.~O. and {Hunstead}, R. and {Intema}, H.~T. and {Marvil}, J. and {Parker}, Q.~A. and {Sirothia}, S.~K. and {Hurley-Walker}, N. and {Bell}, M. and {Bernardi}, G. and {Bowman}, J.~D. and {Briggs}, F. and {Cappallo}, R.~J. and {Callingham}, J.~R. and {Deshpande}, A.~A. and {Dwarakanath}, K.~S. and {For}, B.-Q. and {Greenhill}, L.~J. and {Hancock}, P. and {Hazelton}, B.~J. and {Hindson}, L. and {Johnston-Hollitt}, M. and {Kapi{\'n}ska}, A.~D. and {Kaplan}, D.~L. and {Lenc}, E. and {Lonsdale}, C.~J. and {McKinley}, B. and {McWhirter}, S.~R. and {Mitchell}, D.~A. and {Morales}, M.~F. and {Morgan}, E. and {Morgan}, J. and {Oberoi}, D. and {Offringa}, A. and {Ord}, S.~M. and {Prabu}, T. and {Procopio}, P. and {Udaya Shankar}, N. and {Srivani}, K.~S. and {Staveley-Smith}, L. and {Subrahmanyan}, R. and {Tingay}, S.~J. and {Wayth}, R.~B. and {Webster}, R.~L. and {Williams}, A. and {Williams}, C.~L. and {Wu}, C. and {Zheng}, Q. and {Bannister}, K.~W. and {Chippendale}, A.~P. and {Harvey-Smith}, L. and {Heywood}, I. and {Indermuehle}, B. and {Popping}, A. and {Sault}, R.~J. and {Whiting}, M.~T.},
        title = "{The radio spectral energy distribution of infrared-faint radio sources}",
      journal = {\aap},
         year = 2016,
        month = oct,
       volume = {593},
          eid = {A130},
        pages = {A130},
          doi = {10.1051/0004-6361/201527000},
archivePrefix = {arXiv},
       eprint = {1607.02707},
 primaryClass = {astro-ph.GA},
       adsurl = {https://ui.adsabs.harvard.edu/abs/2016A&A...593A.130H}
}

@ARTICLE{GordonCSS,
       author = {{Gordon}, Yjan A. and {O'Dea}, Christopher P. and {Baum}, Stefi A. and {Bechtol}, Keith and {Duggal}, Chetna and {Ferguson}, Peter S.},
        title = "{Compact Steep Spectrum Radio Sources with Enhanced Star Formation Are Smaller Than 10 kpc}",
      journal = {\apjl},
         year = 2023,
        month = may,
       volume = {948},
       number = {1},
          eid = {L9},
        pages = {L9},
          doi = {10.3847/2041-8213/accf0a},
archivePrefix = {arXiv},
       eprint = {2304.10538},
 primaryClass = {astro-ph.GA},
       adsurl = {https://ui.adsabs.harvard.edu/abs/2023ApJ...948L...9G}
}

@INPROCEEDINGS{FIRST,
       author = {{Becker}, Robert H. and {White}, Richard L. and {Helfand}, David J.},
        title = "{The VLA's FIRST Survey}",
    booktitle = {Astronomical Data Analysis Software and Systems III},
         year = 1994,
       editor = {{Crabtree}, D.~R. and {Hanisch}, R.~J. and {Barnes}, J.},
       series = {Astronomical Society of the Pacific Conference Series},
       volume = {61},
        month = jan,
        pages = {165},
       adsurl = {https://ui.adsabs.harvard.edu/abs/1994ASPC...61..165B}
}

@ARTICLE{VLASS,
       author = {{Lacy}, M. and {Baum}, S.~A. and {Chandler}, C.~J. and {Chatterjee}, S. and {Clarke}, T.~E. and {Deustua}, S. and {English}, J. and {Farnes}, J. and {Gaensler}, B.~M. and {Gugliucci}, N. and {Hallinan}, G. and {Kent}, B.~R. and {Kimball}, A. and {Law}, C.~J. and {Lazio}, T.~J.~W. and {Marvil}, J. and {Mao}, S.~A. and {Medlin}, D. and {Mooley}, K. and {Murphy}, E.~J. and {Myers}, S. and {Osten}, R. and {Richards}, G.~T. and {Rosolowsky}, E. and {Rudnick}, L. and {Schinzel}, F. and {Sivakoff}, G.~R. and {Sjouwerman}, L.~O. and {Taylor}, R. and {White}, R.~L. and {Wrobel}, J. and {Andernach}, H. and {Beasley}, A.~J. and {Berger}, E. and {Bhatnager}, S. and {Birkinshaw}, M. and {Bower}, G.~C. and {Brandt}, W.~N. and {Brown}, S. and {Burke-Spolaor}, S. and {Butler}, B.~J. and {Comerford}, J. and {Demorest}, P.~B. and {Fu}, H. and {Giacintucci}, S. and {Golap}, K. and {G{\"u}th}, T. and {Hales}, C.~A. and {Hiriart}, R. and {Hodge}, J. and {Horesh}, A. and {Ivezi{\'c}}, {\v{Z}}. and {Jarvis}, M.~J. and {Kamble}, A. and {Kassim}, N. and {Liu}, X. and {Loinard}, L. and {Lyons}, D.~K. and {Masters}, J. and {Mezcua}, M. and {Moellenbrock}, G.~A. and {Mroczkowski}, T. and {Nyland}, K. and {O'Dea}, C.~P. and {O'Sullivan}, S.~P. and {Peters}, W.~M. and {Radford}, K. and {Rao}, U. and {Robnett}, J. and {Salcido}, J. and {Shen}, Y. and {Sobotka}, A. and {Witz}, S. and {Vaccari}, M. and {van Weeren}, R.~J. and {Vargas}, A. and {Williams}, P.~K.~G. and {Yoon}, I.},
        title = "{The Karl G. Jansky Very Large Array Sky Survey (VLASS). Science Case and Survey Design}",
      journal = {\pasp},
         year = 2020,
        month = mar,
       volume = {132},
       number = {1009},
          eid = {035001},
        pages = {035001},
          doi = {10.1088/1538-3873/ab63eb},
archivePrefix = {arXiv},
       eprint = {1907.01981},
 primaryClass = {astro-ph.IM},
       adsurl = {https://ui.adsabs.harvard.edu/abs/2020PASP..132c5001L}
}

\newpage
\onecolumn
\appendix 

\section{Tables}
The characterised parameters in this work for all the sources are presented in tables below, separately for the GP and oGP samples.
\subsection{GP sample}

\begin{table*}
    \centering
    \setlength{\tabcolsep}{12pt}

    \begin{tabular}{|l|l|l|l|l|l|}
    \hline
        S-name & $\alpha^{\mathrm{NVSS}}_{\mathrm{TGSS}}$ & $r$ & $\alpha$ & $\chi_{r}^2$ & classification \\ \hline
        
        052801+372826 & -1.52 & 1.25 ± 0.23 & -1.4 ± 0.09 & 0.36 & compact \\ \hline
        053855+382252 & -1.73 & 0.51 ± 0.12 & -1.44 ± 0.09 & 0.28 & non compact \\ \hline
        053959+240608 & -1.87 & 0.96 ± 0.03 & -0.97 ± 0.08 & 1.12 & compact \\ \hline
        054901+274939 & -1.64 & 0.51 ± 0.12 & -1.4 ± 0.1 & 0.14 & non compact \\ \hline
        061232+125158 & -1.58 & 0.82 ± 0.08 & -1.03 ± 0.08 & 0.1 & non compact \\ \hline
        062115+204621 & -1.58 & 0.89 ± 0.23 & -1.47 ± 0.09 & 0.93 & compact \\ \hline
        062726+045759 & -2.06 & 0.76 ± 0.1 & -1.77 ± 0.06 & 1.64 & non compact, ultra-steep \\ \hline
        062844+051917 & -1.81 & 0.77 ± 0.08 & -1.41 ± 0.06 & 0.22 & non compact \\ \hline
        063149+102417 & -1.9 & 0.81 ± 0.09 & -1.36 ± 0.06 & 0.66 & compact \\ \hline
        063653+092618 & -1.56 & 1.19 ± 0.36 & -1.35 ± 0.09 & 0.11 & compact \\ \hline
        063739+093801 & -1.74 & -  & -1.24 ± 0.09 & 0.04 & diffuse \\ \hline
        065642-090241 & -1.5 & 0.91 ± 0.14 & -1.38 ± 0.08 & 0.25 & compact \\ \hline
        070534-052528 & -1.58 & 1.02 ± 0.38 & -1.81 ± 0.11 & 0.74 & compact, ultra-steep \\ \hline
        070800-022931 & -1.67 & 0.43 ± 0.14 & -1.47 ± 0.11 & 0.04 & non compact \\ \hline
        070818-050739 & -1.55 & 0.87 ± 0.17 & -1.44 ± 0.07 & 1.12 & compact \\ \hline
        073245-220852 & -1.6 & - & -1.3 ± 0.09 & 4.21 & resolved \\ \hline
        074705-284142 & -1.62 & 0.67 ± 0.14 & -1.42 ± 0.09 & 1.11 & non compact \\ \hline
        074712-173357 & -1.79 & 0.6 ± 0.09 & -1.4 ± 0.07 & 3.2 & non compact \\ \hline
        080004-280411 & -1.51 & 0.39 ± 0.07 & -1.03 ± 0.09 & 0.00003 & non compact \\ \hline
        084646-375152 & -1.52 & - & -1.05 ± 0.06 & 3.45 & resolved  \\ \hline
        165917-394001 & -1.52 & - & -1.05 ± 0.04 & 0.42 & resolved \\ \hline
        170127-400417 & -1.81 & 0.82 ± 0.05 & -0.87 ± 0.05 & 1.63 & non compact \\ \hline
        171331-381603 & -1.75 & 0.67 ± 0.01 & -0.65 ± 0.05 & 2.05 & non compact \\ \hline
        171360-325254 & -1.81 & 0.27 ± 0.09 & -1.31 ± 0.09 & 0.3 & non compact \\ \hline
        171812-365627 & -1.77 & 0.91 ± 0.01 & -0.93 ± 0.05 & 0.69 & compact \\ \hline
        171929-340058 & -1.83 & 0.73 ± 0.05 & -1.17 ± 0.05 & 0.89 & non compact \\ \hline
        172510-384135 & -1.58 & 0.73 ± 0.02 & -1.11 ± 0.04 & 0.0018 & non compact \\ \hline
        172807-383110 & -1.7 & - & -1.11 ± 0.04 & 2.07 & resolved  \\ \hline
        173154-372155 & -2.03 & - & -1.35 ± 0.08 & 1.47 & resolved \\ \hline
        173324-312616 & -1.77& - & -0.98 ± 0.07 & 2.78 & resolved  \\ \hline
        173728-285519 & -1.83 & - & -1.03 ± 0.06 & 0.56 & resolved \\ \hline
        174028-304357 & -1.65 & 0.88 ± 0.02 & -0.87 ± 0.04 & 0.53 & compact \\ \hline
        174254-260721 & -1.77 & 0.9 ± 0.09 & -1.34 ± 0.09 & 0.08 & compact \\ \hline
        174407-312114 & -1.7 & 1 ± 0.06 & -1.06 ± 0.07 & 0.76 & compact \\ \hline
        174700-350555 & -1.65 & 1.3 ± 0.46 & -2.4 ± 0.12 & - & compact, ultra-steep   \\ \hline

\end{tabular}
\caption{The GP sample is listed with source names given in J2000 RA–DEC, along with their original spectral index upper limits (second column). The measured compactness, $r$, is provided in the third column, and the newly calculated spectral indices, $\alpha$, in the fourth column. The reduced chi-square statistic, $\chi_{r}^2$, is reported for sources detected at all three frequencies (TGSS, RACS-low, and RACS-mid), indicating that three-point flux density measurements were used to fit $\alpha$. Sources without this statistic correspond either to \textit{ND} classifications or to cases lacking a detection at RACS-mid (i.e., where $\alpha$ is derived from two-point flux measurements). The final column lists the resulting classification.}
\label{tab:table1}
\end{table*}

\begin{table*}
    
    \centering
    \setlength{\tabcolsep}{12pt}
    \begin{tabular}{|l|l|l|l|l|l|}

        S-name & $\alpha^{\mathrm{NVSS}}_{\mathrm{TGSS}}$ & $r$ & $\alpha$ & $\chi_{r}^2$ & classification \\ \hline

174926-263840 & -1.79 & - & -0.73 ± 0.09 & 3.71 & resolved  \\ \hline
175003-274816 & -1.52 & 0.72 ± 0.03 & -0.71 ± 0.08 & 0.62 & non compact \\ \hline
175041-254603 & -1.61 & 0.82 ± 0.06 & -1.06 ± 0.07 & 0.5 & non compact \\ \hline
175113-273724 & -2.12 & 0.65 ± 0.18 & -2.09 ± 0.14 & -  & non compact, ultra-steep  \\ \hline  
175520-255704 & -1.53 & - & -0.57 ± 0.07 & 4.61 & resolved \\ \hline
175956-295540 & -1.61 & 0.99 ± 0.28 & -1.28 ± 0.09 & 0.22 & compact \\ \hline
180156-244858 & -1.8 & 0.89 ± 0.03 & -0.99 ± 0.05 & 0.77 & compact \\ \hline
180209-250434 & -1.93 & - & -0.7 ± 0.05 & 0.1 & resolved \\ \hline
180224-233536 & -1.66 & - & -0.63 ± 0.08 & 0.1 & resolved \\ \hline
180245-251613 & -1.8 & 0.95 ± 0.03 & -1.05 ± 0.07 & 0.3 & compact \\ \hline
180343-235255 & -1.62 & - & -0.81 ± 0.05 & 4.13 & resolved \\ \hline
180939-223114 & -1.64 & 0.77 ± 0.1 & -1.36 ± 0.1 & 0.33 & non compact \\ \hline
180943-194913 & -1.91 & 0.92 ± 0.03 & -1.13 ± 0.04 & 0.38 & compact \\ \hline
181343-170627 & -1.66 & 0.9 ± 0.02 & -0.71 ± 0.06 & 3.8 & compact \\ \hline
181449-123901 & -1.69 & 1.07 ± 0.1 & -1.12 ± 0.09 & 5.8 & compact \\ \hline
181742-173158 & -1.68 & 0.95 ± 0.03 & -0.78 ± 0.05 & 1.65 & compact \\ \hline
181823-181327 & -1.57 & 0.59 ± 0.03 & -0.67 ± 0.07 & 0.71 & non compact \\ \hline
182249-203940 & -1.88 & 0.99 ± 0.37 & -1.63 ± 0.08 & 0.49 & compact \\ \hline
182736-084941 & -2.13 & - & < -2.55 & - & ND, ultra-steep \\ \hline
182748-130153 & -1.58 & 0.94 ± 0.02 & -0.86 ± 0.05 & 2.27 & compact \\ \hline
183113-021012 & -1.81 & 0.93 ± 0.01 & -0.94 ± 0.03 & 0.04 & compact \\ \hline
183227-090420 & -1.58 & 0.76 ± 0.02 & -0.86 ± 0.07 & 3.91 & non compact \\ \hline
183545-142139 & -1.5 & - & < -1.99 & - & ND, ultra-steep \\ \hline
183934-123329 & -1.68 & - & < -2.13 & - & ND, ultra-steep \\ \hline
184046+021858 & -1.75 & 0.7 ± 0.22 & -1.83 ± 0.08 & 0.35 & compact, ultra-steep \\ \hline
184404-013650 & -1.74 & 0.86 ± 0.03 & -0.78 ± 0.07 & 0.2 & non compact \\ \hline
184631+003604 & -1.79 & - & -1.49 ± 0.11 & 0.07 & resolved \\ \hline
185445+005811 & -1.57 & - & -0.84 ± 0.07 & 0.19 & resolved \\ \hline
185940+061154 & -1.79 & 0.34 ± 0.08 & -1.35 ± 0.1 & 8.75 & non compact \\ \hline
190042+085919 & -1.77 & 1.11 ± 0.26 & -1.96 ± 0.09 & 0.56 & compact, ultra-steep \\ \hline
190104+030141 & -1.62 & 0.68 ± 0.03 & -0.9 ± 0.08 & 0.63 & non compact \\ \hline
190134-012527 & -2.36 & - & < -3.12 & - & ND, ultra-steep \\ \hline
190749+094235 & -1.71 & - & -0.7 ± 0.08 & 7.44 & resolved \\ \hline
191131+094320 & -1.81 & 0.97 ± 0.04 & -0.94 ± 0.05 & 2.23 & compact \\ \hline
192442+202721 & -1.75 & - & < -2.44 & - & ND, ultra-steep \\ \hline
192717+143904 & -1.54 & - & -0.88 ± 0.11 & 0.89 & resolved \\ \hline
192944+155006 & -1.91 & 0.96 ± 0.04 & -1.07 ± 0.06 & 0.000117 & compact \\ \hline
194819+221259 & -1.71 & 0.75 ± 0.13 & -1.55 ± 0.09 & 0.3 & non compact \\ \hline
195952+335245 & -1.56 & 1.08 ± 0.19 & -1.43 ± 0.08 & 0.87 & compact \\ \hline
201760+363018 & -1.74 & 1.02 ± 0.03 & -0.87 ± 0.08 & 1.77 & compact \\ \hline
203020+382337 & -1.64 & 0.9 ± 0.05 & -1.1 ± 0.09 & 0.82 & compact \\ \hline
203401+401009 & -2.23 & 0.78 ± 0.01 & -1.04 ± 0.04 & 1.5 & non compact \\ \hline
203759+365249 & -1.52 & 0.86 ± 0.08 & -1.2 ± 0.1 & 1.06 & compact \\ \hline

\end{tabular}
\end{table*}

\clearpage

\subsection{oGP sample}

\begin{table*}
    \centering
    \setlength{\tabcolsep}{12pt}
    \begin{tabular}{|l|l|l|l|l|l|}
    \hline
        S-name & $\alpha^{\mathrm{NVSS}}_{\mathrm{TGSS}}$ & $r$ & $\alpha$ & $\chi_{r}^2$ & classification \\ \hline
        
        000533+130309 & -1.87 & 0.89 ± 0.08 & -1.4 ± 0.05 & 0.77 & compact \\ \hline
001639-124203 & -2.22 & 0.64 ± 0.05 & -1.41 ± 0.04 & 5.52 & non compact \\ \hline
001806-201542 & -2.03 & 0.36 ± 0.14 & -1.86 ± 0.1 & 0.13 & non compact, ultra-steep \\ \hline
001857-122312 & -1.95 & 0.59 ± 0.04 & -1.01 ± 0.05 & 1.41 & non compact \\ \hline
011230+000103 & -1.83 & 0.75 ± 0.22 & -2.03 ± 0.08 & 0.12 & compact, ultra-steep \\ \hline
014329-243214 & -2.23 & 0.86 ± 0.14 & -1.92 ± 0.05 & 2.11 & compact, ultra-steep \\ \hline
020641-175934 & -1.82 & 0.43 ± 0.09 & -1.4 ± 0.07 & 1.67 & non compact \\ \hline
021223+322625 & -2.1 & - & -1.55 ± 0.05 & 1.09 & resolved \\ \hline
022546-263559 & -2.06 & 0.57 ± 0.25 & -2.04 ± 0.11 & 0.21 & non compact, ultra-steep \\ \hline
023045-313221 & -1.83 & - & -1.43 ± 0.06 & 2.28 & resolved \\ \hline
023640+115015 & -1.89 & 0.83 ± 0.13 & -1.63 ± 0.06 & 6.18 & compact \\ \hline
023821-261558 & -1.98 & 0.27 ± 0.06 & -1.5 ± 0.07 & 0.08 & non compact \\ \hline
023837-225420 & -1.87 & 0.45 ± 0.05 & -1.13 ± 0.07 & 1.25 & non compact \\ \hline
025839+055150 & -1.87 & 0.85 ± 0.1 & -1.55 ± 0.06 & 1 & compact \\ \hline
030427+121326 & -1.82 & 1.19 ± 0.24 & -1.79 ± 0.07 & 0.05 & compact, ultra-steep \\ \hline
032203-373638 & -1.91 & - & -1.2 ± 0.04 & 8.66 & resolved \\ \hline
035628+164134 & -1.82 & 0.43 ± 0.11 & -1.44 ± 0.09 & 2.37 & non compact \\ \hline
040407-110730 & -1.98 & 0.64 ± 0.23 & -2.05 ± 0.11 & 0.03 & non compact, ultra-steep \\ \hline
041223-005620 & -2.12 & - & < -2.58 & - & ND, ultra-steep \\ \hline
041223-010146 & -2.07 & - & < -2.46 & - & ND, ultra-steep \\ \hline
050159+164018 & -1.8 & 0.88 ± 0.4 & -2.07 ± 0.14 & - & compact, ultra-steep \\ \hline
050612+154100 & -1.83 & 1.12 ± 0.15 & -1.56 ± 0.06 & 0.01 & compact \\ \hline
050922+085625 & -1.82 & 0.78 ± 0.16 & -1.67 ± 0.06 & 5.84 & compact \\ \hline
051026+022605 & -1.93 & 0.59 ± 0.15 & -1.45 ± 0.07 & 2.27 & non compact \\ \hline
054110+201619 & -1.85 & 0.91 ± 0.05 & -1.12 ± 0.08 & 0.34 & compact \\ \hline
054719+384106 & -1.8 & - & -1.41 ± 0.09 & 7.97 & resolved \\ \hline
054916+140130 & -1.82 & 0.49 ± 0.1 & -1.65 ± 0.1 & - & non compact \\ \hline
061527-221846 & -1.97 & - & -1.94 ± 0.11 & 0.06 & diffuse, ultra-steep \\ \hline
063119-360457 & -1.9 & - & -1.47 ± 0.06 & 7.92 & resolved \\ \hline
065048+134441 & -1.8 & 0.58 ± 0.11 & -1.53 ± 0.08 & 1.15 & non compact \\ \hline
071512+245001 & -2.01 & 0.94 ± 0.09 & -1.54 ± 0.05 & 0.02 & compact \\ \hline
072649-385057 & -1.83 & 0.99 ± 0.08 & -1.44 ± 0.05 & 0.78 & compact \\ \hline
072657+220801 & -1.85 & - & -1.29 ± 0.07 & 0.00376 & resolved \\ \hline
075537-194227 & -2.02 & - & -1.6 ± 0.06 & 1.23 & resolved \\ \hline
082712-311201 & -2.11 & 0.89 ± 0.13 & -1.95 ± 0.05 & 1.3 & compact, ultra-steep \\ \hline
083639+210159 & -1.83 & 0.63 ± 0.218 & -1.61 ± 0.09 & 0.03 & non compact \\ \hline
084451+180342 & -1.83 & 0.61 ± 0.23 & -1.9 ± 0.11 & 0.0008 & non compact, ultra-steep \\ \hline

\end{tabular}
    \caption{The oGP sample, with the source names as J2000 RA-DEC, their original spectral index upper limits (second column), compactness $r$, calculated $\alpha$, goodness of fit $\chi_{r}^2$, and the final classification.}
    \label{tab:table2}
\end{table*}

\begin{table*}
    
    \centering
    \setlength{\tabcolsep}{12pt}
    \begin{tabular}{|l|l|l|l|l|l|}
        S-name & $\alpha^{\mathrm{NVSS}}_{\mathrm{TGSS}}$ & $r$ & $\alpha$ & $\chi_{r}^2$ & classification \\ \hline

090022-091723 & -1.86 & 0.76 ± 0.24 & -2.13 ± 0.1 & - & compact, ultra-steep \\ \hline
090727-133914 & -2.01 & - & < -2.59 & - & ND, ultra-steep \\ \hline
091211-165016 & -2.02 & - & -1.59 ± 0.07 & 2.55 & resolved \\ \hline
092210-142628 & -2.2 & - & < -2.61 & - & ND, ultra-steep \\ \hline
094019+364153 & -2.06 & 0.87 ± 0.16 & -1.89 ± 0.05 & 0.48 & compact, ultra-steep \\ \hline
100638-350925 & -1.88 & 0.44 ± 0.14 & -1.59 ± 0.08 & 2.01 & non compact \\ \hline
103156-233812 & -1.96 & 0.46 ± 0.07 & -1.46 ± 0.06 & 0.76 & non compact \\ \hline 
103546-232323 & -1.9 & - & -1.64 ± 0.08 & 9.96 & resolved \\ \hline
104510-311110 & -1.81 & - & -1.43 ± 0.07 & 2.1 & diffuse \\ \hline
104539-313818 & -1.8 & 0.72 ± 0.16 & -2.03 ± 0.11 & - & non compact, ultra-steep \\ \hline
105207+072207 & -1.84 & - & -1.49 ± 0.09 & 0.24 & diffuse \\ \hline
110416+350129 & -1.87 & - & -1.57 ± 0.09 & 1.96 & resolved \\ \hline
110639-211248 & -2.17 & - & < -2.5 & - & ND, ultra-steep \\ \hline
110854-245959 & -1.8 & 0.31 ± 0.08 & -1.43 ± 0.08 & 1.01 & non compact \\ \hline
111526-165143 & -1.84 & - & < 2.32 & - & ND, ultra-steep \\ \hline
113439-173126 & -2.03 & - & < -2.59 & - & ND, ultra-steep \\ \hline
113846-132448 & -2.19 & 0.75 ± 0.1 & -1.52 ± 0.05 & 1.7 & non compact \\ \hline
114226-224418 & -1.94 & 0.78 ± 0.21 & -1.99 ± 0.07 & 3.27 & compact, ultra-steep \\ \hline
121424-255200 & -1.84 & 1.14 ± 0.1 & -1.34 ± 0.05 & 0.96 & compact \\ \hline
123038+324721 & -1.82 & - & -2.1 ± 0.15 & 0.01 & diffuse, ultra-steep \\ \hline
133708-103232 & -1.85 & 0.91 ± 0.09 & -1.59 ± 0.05 & 1.52 & compact \\ \hline
135729-084653 & -1.88 & 0.72 ± 0.27 & -2.34 ± 0.12 & - & compact, ultra-steep \\ \hline
143819+030210 & -1.87 & - & -1.39 ± 0.16 & - & resolved \\ \hline
145549-111208 & -1.84 & - & < -2.15 & - & ND, ultra-steep \\ \hline
154804-325426 & -2 & 0.97 ± 0.09 & -1.63 ± 0.05 & 0.88 & compact \\ \hline
161259+224647 & -1.9 & 0.57 ± 0.11 & -1.49 ± 0.08 & 2.03 & non compact \\ \hline
162206-333608 & -1.98 & 0.97 ± 0.09 & -1.55 ± 0.04 & 1.5 & compact \\ \hline
164159-115641 & -1.84 & 1.14 ± 0.2 & -1.74 ± 0.06 & 0.91 & compact, ultra-steep \\ \hline
164347+342413 & -2 & 0.87 ± 0.16 & -1.78 ± 0.05 & 2.61 & compact, ultra-steep \\ \hline
172059+295453 & -1.9 & 1.11 ± 0.32 & -1.93 ± 0.07 & 0.57 & compact, ultra-steep \\ \hline
172759-160910 & -1.91 & - & -2.46 ± 0.18 & - & diffuse, ultra-steep \\ \hline
173556+063715 & -1.83 & 0.37 ± 0.11 & -1.73 ± 0.1 & 0.62 & non compact, ultra-steep \\ \hline
174010+060630 & -2.14 & - & -1.56 ± 0.06 & 0.005 & resolved \\ \hline
175419-084135 & -1.83 & - & < -2.36 & - & ND, ultra-steep \\ \hline
175910-075925 & -1.99 & - & < -2.8 & - & ND, ultra-steep \\ \hline
180306-054955 & -1.81 & 0.7 ± 0.07 & -1.38 ± 0.08 & 0.69 & non compact \\ \hline
182506+124616 & -1.84 & - & < -2.4 & - & ND, ultra-steep \\ \hline
184009+110207 & -1.84 & 0.97 ± 0.25 & -2.39 ± 0.1 & - & compact, ultra-steep \\ \hline
184757+095010 & -1.88 & 0.87 ± 0.22 & -1.7 ± 0.08 & 1.53 & compact \\ \hline
184816+382637 & -2.11 & 1.09 ± 0.19 & -1.86 ± 0.05 & 1.65 & compact, ultra-steep \\ \hline
180309-353216 & -2.2 & 0.58 ± 0.17 & -2.14 ± 0.08 & 0.53 & non compact, ultra-steep \\ \hline
181403-391903 & -2.08 & 0.7 ± 0.26 & -1.9 ± 0.07 & 1.14 & compact, ultra-steep \\ \hline
181755-395157 & -1.8 & 1.02 ± 0.13 & -1.62 ± 0.06 & 0.03 & compact \\ \hline

\end{tabular}
\end{table*}

\clearpage
\begin{table*}
    
    \centering
    \setlength{\tabcolsep}{12pt}
    \begin{tabular}{|l|l|l|l|l|l|}
        S-name & $\alpha^{\mathrm{NVSS}}_{\mathrm{TGSS}}$ & $r$ & $\alpha$ & $\chi_{r}^2$ & classification \\ \hline

182432-245212 & -2.46 & 0.55 ± 0.13 & -2.21 ± 0.06 & 1.19 & non compact, ultra-steep \\ \hline
192549-293352 & -1.95 & 0.78 ± 0.07 & -1.12 ± 0.06 & 3.17 & non compact \\ \hline
194557-174125 & -1.82 & - & -1.74 ± 0.12 & 0.07 & diffuse, ultra-steep \\ \hline
195902-141043 & -1.84 & 0.46 ± 0.17 & -1.47 ± 0.11 & 0.02 & non compact \\ \hline
204908+322547 & -1.87 & 0.93 ± 0.18 & -1.69 ± 0.06 & 1.81 & compact \\ \hline
211654-205319 & -1.9 & - & < -2.17 & - & ND, ultra-steep \\ \hline
212255+001152 & -2.24 & 0.4 ± 0.09 & -1.82 ± 0.06 & 4.33 & non compact, ultra-steep \\ \hline
220748-232512 & -1.84 & 0.98 ± 0.09 & -1.23 ± 0.05 & 2.76 & compact \\ \hline
222232-093222 & -2.01 & - & < -2.73 & - & ND, ultra-steep \\ \hline
223441-175227 & -1.88 & 1.29 ± 0.2 & -1.64 ± 0.06 & 0.96 & compact \\ \hline
233043+195344 & -1.8 & 1.03 ± 0.15 & -1.58 ± 0.06 & 1.37 & compact \\ \hline
235022-352643 & -2.03 & 0.84 ± 0.08 & -1.64 ± 0.04 & 0.9 & compact \\ \hline
235931-343900 & -1.83 & 0.95 ± 0.2 & -1.43 ± 0.07 & 1.64 & compact \\ \hline

    \end{tabular}

\end{table*}



\label{lastpage}

\end{document}